\documentclass[10pt,twocolumn,twoside]{IEEEtran} 
\usepackage{graphics} % for pdf, bitmapped graphics files
\usepackage{epsfig} % for postscript graphics files
\usepackage{mathptmx} % assumes new font selection scheme installed
\usepackage{times} % assumes new font selection scheme installed
\usepackage{amsmath} % assumes amsmath package installed
\usepackage{amssymb}  % assumes amsmath package installed
\usepackage{cite}
\usepackage{graphicx} 
\usepackage{tabularx}
\usepackage{hyperref}
\usepackage{xcolor}
\usepackage{multicol}
\usepackage{bbm} 
\usepackage{dsfont}
\usepackage{mathtools}

\newtheorem{thm}{Theorem}
\newtheorem{lem}{Lemma}
\newtheorem{prop}{Proposition}
\newtheorem{rem}{Remark}
\newtheorem{defi}{Definition}

\DeclareMathOperator*{\esssup}{ess \, sup}

\newcommand{\dx}{\, \mathrm d x}
\newcommand{\dy}{\, \mathrm d y}
\newcommand{\R}{\mathbb R}
\newcommand{\eps}{\varepsilon}

\newcommand{\ind}{\mathds{1}}
\newcommand{\dxi}{\, \mathrm d \xi}
\newcommand{\E}{\mathrm{E}}

\NewDocumentCommand{\vertiii}{sO{}m}{%
  \IfBooleanTF{#1}{% automatic scaling, use with care
    \left|\opnormkern\left|\opnormkern\left|
    #3
    \right|\opnormkern\right|\opnormkern\right|
  }{
    \mathopen{#2|\opnormkern #2|\opnormkern #2|}
    #3
    \mathclose{#2|\opnormkern #2|\opnormkern #2|}
  }%
}
\newcommand{\opnormkern}{\mkern-1.5mu\relax}

\usepackage{comment}

\title{\LARGE \bf Opinion dynamics on signed graphs and graphons
}

\author{Raoul Prisant, Federica Garin, Paolo Frasca  
\thanks{All authors are with Univ.\ Grenoble Alpes, CNRS, Inria, Grenoble INP, GIPSA-lab, 38000 Grenoble, France (e-mails: raoul.prisant@gipsa-lab.fr; federica.garin@inria.fr; paolo.frasca@gipsa-lab.fr).} 
\thanks{This work has been partly supported by the French National Research Agency through grant COCOON ANR-22-CE48-0011.}
}

\begin{document}

\maketitle

\thispagestyle{empty}
\pagestyle{empty}

%%%%%%%%%%%%%%%%%%%%%%%%%%%%%%%%%%%%%%%%%%%%%%%%%%%%%%%%%%%%%%%%%%%%%%%%%%%%%%%%
\begin{abstract}
    In this paper, we make use of graphon theory to study opinion dynamics on large undirected networks. The opinion dynamics models that we take into consideration allow for negative interactions between the individuals, whose opinions can thus grow apart. We consider both the repelling and the opposing models of negative interactions, which have been studied in the literature. We define the repelling and the opposing dynamics on signed graphons and we show that their initial value problem solutions exist and are unique. We then show that, in a suitable sense, the graphon dynamics is a good approximation of the dynamics on large graphs that converge to a graphon. This result applies to large random graphs that are sampled according to a graphon (W-random graphs), for which we provide a new convergence result under very general assumptions; in particular, the graphs' average degrees only need to grow faster than $\log n$.
\end{abstract}

\begin{IEEEkeywords}
 Graphons; W-random graphs; graph Laplacian; signed graphs; opinion dynamics; social networks
\end{IEEEkeywords}

\section{Introduction}
The control systems community has developed a keen interest in the analysis of mathematical models of opinion dynamics on social networks. 
A key question in this research, which has been summarized by several survey papers~\cite{proskurnikov2017tutorial,proskurnikov2018tutorial,shi2019dynamics,tian2023dynamics,bernardo2024bounded},
has been to determine whether or not the individuals in the social network reach a consensus, whereby their opinions are in agreement~\cite{friedkin2015problem}. A popular feature of non-agreement models is the presence of negative (i.e.\ antagonistic) interactions between individuals~\cite{altafini2012consensus,shi2019dynamics}: indeed, negative interactions counter the positive interactions that are commonplace in social influence models and  drive individuals to conform with their peers.

Opinion dynamics take place on social networks, which can be very large, and their analysis needs to cope with such scale. A useful instrument in this perspective is the theory of graph limits, and particularly of graphons~\cite{lovasz2012book}. Using graphons to study dynamical systems on networks is currently a very active area of research. %, which extends beyond control systems research~\cite{bonnet2022consensus,}.
Some works have focused on relevant properties of the networks~\cite{avella2018centrality,vizuete2021laplacian,belabbas2021h}, while others have considered dynamical systems on graphons, including random walks~\cite{petit2021random}, interacting particle systems~\cite{allmeier2024accuracy}, 
epidemics~\cite{delmas2023individual,erol2023contagion}, opinion dynamics~\cite{bonnet2022consensus,cdc2024}, synchronization~\cite{medvedev2019,nagpal2024synchronization}, and general control systems~\cite{gao2017control,gao2019graphon,gao2021subspace}.

In this paper we consider opinion dynamics on signed graphs and we define their graphon counterparts. More precisely, we consider both the so-called {\em repelling model} and the {\em opposing model}, two  extensions of the classical French-DeGroot model that allow the agents to have negative interactions~\cite{shi2019dynamics}. 
After showing, in Theorem~\ref{thm:existence-and-unicity}, the existence and uniqueness of the solutions to these two dynamics on signed graphons, we offer two main contributions. 
First, in Theorem~\ref{thm:bound} we estimate the approximation error between solutions on signed graphs and solutions on signed graphons.
The error bound in Theorem~\ref{thm:bound} readily implies the convergence of the solutions on signed graphs to solutions on a signed graphon, provided the sequence of graphs converges to the graphon as the number of individuals $n$ goes to infinity. 
Second, in Theorem~\ref{thm:sampled-convergence} we prove that these sufficient conditions are general enough to apply to sequences of random signed graphs sampled from signed graphons. Our analysis distinguishes itself from the literature by making the weakest assumptions on the regularity of the graphons and by allowing for {\em sparse} sampled graphs, whose average degree only needs to grow faster than $\log n$. These features are enabled by two results of independent interest: Theorem~\ref{convergence_sampled_to_graphon_NEW}, which contains new convergence conditions for sparse graphs that are sampled according to graphons (also known as W-random graphs), and Theorem~\ref{thm:degrees}, which provides tight estimates of the maximum and the average degrees of these
sampled graphs.

\paragraph*{Related work} 
Several papers have considered opinion dynamics on graphons, related Laplacian-based dynamics on graphons, or the properties of graphon Laplacians.
The overwhelming majority of works assume
non-negative interactions, including~\cite{vonluxburg2008consistency,vizuete2021laplacian,bramburger2023pattern,petit2021random}.
Papers~\cite{bonnet2022consensus,ayi2021mean} contain a study of graphon opinion dynamics with non-negative, but possibly time-varying, interactions. 
Other works have considered more general, nonlinear, Laplacian-based dynamics on networks. For instance,  the well-known work by Medvedev~\cite{medvedev2014nonlinear,medvedev2014Wrandom}
covers nonlinear dynamics such as the Kuramoto model of synchronization. 
Signed graphons have been sometimes considered in the literature~\cite{lovasz2011subgraph}, but rarely as a dynamical model, even though the importance of large-scale signed social networks has long been recognized~\cite{facchetti2011computing}.
Recently, the paper \cite{aletti2022opinion} has considered the repelling dynamics on signed graphs and has proved the existence of solutions for piece-wise constant graphons. 
Our results are more general than the latter as they cover the opposing dynamics, include explicit approximation bounds, consider the case of random sampled graphs, and make much more general assumptions on the regularity of the graphons. 
A more detailed discussion of our contributions compared to prior work is deferred to Section~\ref{sect:discussion}.

\paragraph*{Outline}
In Section~\ref{sect:opinion-dynamics}, we recall the relevant models of opinion dynamics on signed graphs and we define their graphon counterparts. We then prove that the latter have complete classical solutions from any initial condition. In Section~\ref{sect:convergence-solutions}, we prove a bound on the error between solutions of the dynamics on graphs with $n$ nodes and solutions of the graphon dynamics. In Section~\ref{sect:convergence-sampled} we show that this bound implies that solutions converge, when $n\to\infty$, for large random graphs sampled from signed graphons. Section~\ref{sect:discussion} contains a detailed comparison with the literature and Section~\ref{sect:simulations} an illustrative example, before concluding remarks in Section~\ref{sect:conclusion}.

\section{Opinion dynamics on graphs and graphons}\label{sect:opinion-dynamics}

In this section, we recall established models of opinion dynamics on signed graphs and we define their counterparts on signed graphons. On our way to the latter, we recall the necessary preliminaries about graphons and about random graphs sampled from graphons.

\subsection{Opinion dynamics on signed graphs}
Opinion dynamics study the evolution of the opinions of interacting individuals~\cite{proskurnikov2017tutorial,proskurnikov2018tutorial}. The interaction network is modeled by an undirected graph $G_n$, where the nodes $v_i$ represent the individuals and the edges $e_{ij}$ represent an interaction between $v_i$ and $v_j$. In general, a weight matrix $A^{(n)}$ 
can be assigned to determine the strength of each connection.
For the scope of this paper, we will take symmetric matrices $A^{(n)}$ with entries $A_{ij}^{(n)}\in\{-1,0,+1\}$, where value $+1$ represents a positive interaction, that is, {the two individuals trust each other and therefore their opinions grow closer, value $0$ means lack of interaction, and value $-1$ represents the interaction between individuals who dislike each other and whose opinions tend to grow apart. 
Here, $A^{(n)}$ is a given matrix, but we highlight the dependence on the size $n$ in the notation, since throughout the paper we will consider sequences of graphs of increasing size.}

Two relevant ways of defining an opinion dynamics associated with such signed graphs have been considered in the literature~\cite[Sect.~2.5.2]{shi2019dynamics}. 
The first is the so-called {\em repelling} model, in which the opinion of each node $i$ is modeled as $u^{(n)}_i\in\mathbb{R}$, and its evolution in time is determined by the differential equation
%\todo{aggiungere i (t), usare le A maiuscole}
\begin{equation}\label{DynGraphRep}
    \begin{cases}
\dot{u}^{(n)}_i(t)=\displaystyle\alpha_n\sum_{j=1}^n   A_{ij}^{(n)}(u^{(n)}_j(t)-u^{(n)}_i(t)), \\ 
        u^{(n)}_i(0)=g_i^{(n)},
    \end{cases}
\end{equation}
where $n$ is the number of nodes and $g_i^{(n)}$ the initial opinion of node $i$. 
The second is the {\em opposing} model, also known as Altafini model, in which the evolution in time of the opinion is instead described by the equation
\begin{equation}\label{DynGraphOpp}
    \begin{cases}
        \dot{u}^{(n)}_i(t)= \displaystyle\alpha_n\left(\sum_{j=1}^n   A_{ij}^{(n)}u^{(n)}_j(t)-\sum_{j=1}^n|A_{ij}^{(n)}|u^{(n)}_i(t)\right), \\ 
        u^{(n)}_i(0)=g_i^{(n)}.
    \end{cases}
\end{equation}

In both dynamics, the positive coefficient $\alpha_n$ is a ``speed parameter'' that we let depend on $n$: although it could be made to disappear by rescaling the time variable, its explicit presence will be useful in order to let $\alpha_n$ go to zero to ``compensate'' for an increased number of potential interactions when $n$ is large.

Following~\cite{shi2019dynamics}, we observe that in both dynamics the positive interactions are consistent with the classical DeGroot's rule of social interactions,
which postulates that the opinions of trustful social members are attractive to each
other~\cite{degroot1974reaching}. Along a negative link, the opposing rule~\cite{altafini2012consensus} postulates that the interaction will drive a node state to
be attracted by the opposite of its neighbor's state; the repelling rule~\cite{shi2013agreement} postulates
that the two node states will repel each other instead of being attractive.

\subsection{Graphs and graphons}
When studying opinion dynamics on a society scale, the dimension of the graph becomes too large to deal with. To address this problem, we will make use of {\em graphons}, which are usually defined as symmetric measurable functions $W:{I}^2\rightarrow [0,1]$, {where $I=[0,1]$}. Throughout this paper, we will define {\em signed graphons} to be symmetric measurable functions $W:{I}^2\rightarrow [-1,1]$. We shall instead refer to bounded symmetric measurable functions taking values in $\R$ simply as kernels. 
We define the degree of a kernel $W$ as the function $d_W:I\to \mathbb{R}^+$ such that $d_W(x)=\int_I W(x,y)\dy.$

For every kernel $W$, 
we can define an integral operator $T_W : L^2{(I)}\rightarrow L^2{(I)}$ by:
\[
\left(T_Wf\right)(x):=\int_{I}W(x,y)f(y) \dy .
\]
Notice that $T_{W_1+W_2}=T_{W_1}+T_{W_2}$ for any two kernels $W_1$, $W_2$.
We will equip the space of kernels with the operator norm defined as
\[
\vertiii{T_W}:=\sup_{\|f\|_2=1}\|T_Wf\|_2.
\]

In order to compare (possibly signed) graphons and graphs of different sizes, it is useful to represent a graph as a piece-wise constant graphon.  To do this, we partition the interval $I$ in $n$ intervals $I_i=(\frac{i-1}{n},\frac{i}{n}]$ for $i=1,\dots,n$, and set {$W_n(x,y)=\sum_{i,j} A^{(n)}_{ij}\mathds{1}_{I_i}(x)\mathds{1}_{I_j}(y)$, that is,}  $W_n(x,y)=A^{(n)}_{ij},$ for all $(x,y)\in I_i\times I_j$. 
Informally, we can then see a graphon as the limit of a sequence of graphs, as for $n\rightarrow\infty$ the interval of length $\frac{1}{n}$ corresponding to each node has size that goes to 0, so that ${I}$ becomes a continuum of nodes.
To make this idea precise, in this paper we will use the distance induced by the operator norm. Hence, we will say that a sequence of graphs (seen as its associated graphons $W_{n}$) converges to a graphon $W$ if $\vertiii{T_{W_n}-T_W}$ goes to zero. {This convergence notion among graphons} is equivalent to the notion induced by the more popular cut norm; see~\cite[Lemma E.6]{janson2013graphons}. Other (non-equivalent) norms are also used in the literature, such as the $L^2$ norm $\|W_n-W\|_2$ that is used in~\cite{medvedev2014nonlinear}.

\subsection{Sampled graphs}
The graph sequences of main interest for this paper are sequences of sampled graphs. The latter are widely studied in the literature for graphons with values in $[0,1]$. For values in $[-1,1]$, the natural extension is the following~\cite{borgs2019Lp}.

\begin{defi}[Sampled signed graphs]
    Given a graphon $W$ and a set of points $X_i\in[0,1]$ with $i=1,\ldots,n$, a sampled signed graph is a random graph with adjacency matrix $A^{(n)}$ such that
    $$
        A^{(n)}_{ij}=\text{sign}(W(X_i,X_j))\cdot \eta_{ij},
    $$
    where $\eta_{ij}\sim Ber(\epsilon_n|W(X_i,X_j)|)$ are independent Bernoulli random variables, for all $i < j$, and $\epsilon_n\in (0,1]$ is a parameter describing the sparsity of the adjacency matrix of the sampled signed graph. \hfill$\square$\label{sampledgraph_negativeweights}
\end{defi}

If $\epsilon_n=1$, when we sample from a graphon $W$ with positive values, we have, for instance, that $W(X_i,X_j)=0.5$ corresponds to a 50\% chance of edge $(i,j)$ existing (i.e. $A^{(n)}_{ij}=1$). The definition is then a natural extension to negative values, where $W(X_i,X_j)=-0.5$ means there is a 50\% chance of a negative edge existing between nodes $i$ and $j$. If $\epsilon_n<1$, the probability of an edge existing between any couple of nodes is reduced, so that the obtained graph is sparser.

When convenient, with a slight abuse of vocabulary, we will call `graph sampled from the graphon $W$' also the piece-wise constant graphon 
$W_n(x,y)=\sum_{i,j} A^{(n)}_{ij}\mathds{1}_{I_i}(x)\mathds{1}_{I_j}(y)$ associated with the actual graph.

For the points $\{X_i\}$, we consider two possible definitions.
\begin{defi}[Latent variables]
For each $n$, consider $n$ points $X_1, \dots, X_n$ in the interval $I$, defined as follows:
\label{latent_var}
    \begin{enumerate}
        \item Deterministic latent variables: $X_i=\frac{i}{n}$, \label{dlv}
        \item Stochastic latent variables: $X_i = U_{(i)}$, where $U_1, \dots, U_n$ are i.i.d.\ uniform in $I$ and $U_{(1)}, \dots, U_{(n)}$ is the corresponding order statistics.\hfill$\square$\label{slv}
    \end{enumerate}
\end{defi}

\subsection{Opinion dynamics on signed graphons}
In analogy to the graph case, for the {\em repelling} model we can define the dynamics on the graphon as 
\begin{equation} \label{DynGraphonRep}
\begin{cases}
    \displaystyle\frac{\partial u}{\partial t}(x,t)=\displaystyle\int_{I}W(x,y)(u(y,t)-u(x,t)) \dy  \\
    u(x,0)=g(x),
\end{cases}
\end{equation}
where {$I=[0,1],$} $u:{I}\times\R^+ \to \R$ is the opinion distribution and $g:{I}\to\R$ is the initial opinion distribution, whereas for the {\em opposing} model, we can write  
\begin{equation}\label{DynGraphonOpp}
\begin{cases}\displaystyle
    \frac{\partial u}{\partial t}(x,t)=\int_{I}W(x,y)u(y,t)\dy   - \int_{I}|W(x,y)|u(x,t) \dy  \\
    u(x,0)=g(x).
\end{cases}
\end{equation}

Existence and uniqueness of solutions of \eqref{DynGraphonRep} is given in
\cite[Theorem~3.2]{medvedev2014nonlinear}; a slight modification of its proof allows to obtain the same for \eqref{DynGraphonOpp} as well.
\begin{thm}[Existence and uniqueness]  \label{thm:existence-and-unicity}
Assume that $W\in L^{\infty}(I^2)$ and $g\in L^{\infty}(I)$. Then,  %for any $T>0$, 
there exists a unique solution of \eqref{DynGraphonRep}, $u\in C^1(\mathbb{R}^+;L^{\infty}(I))$.
The same holds for \eqref{DynGraphonOpp}.%\hfill$\square$
\end{thm}
\begin{IEEEproof}
The statement concerning \eqref{DynGraphonRep} is \cite[Theorem~3.2]{medvedev2014nonlinear}. The proof for \eqref{DynGraphonOpp} can be obtained by a very similar contraction argument, which we report for completeness.
First, notice that $|W(x,y)| =  W(x,y) + 2 W^-(x,y)$, where
$W^-(x,y) = \max\{-W(x,y),0)\}$ is the negative part of $W$.
With this, we can see that the integro-differential equation in \eqref{DynGraphonOpp} is equivalent to the integral equation $u = Ku$ with $K$ defined as 
\begin{multline*}
  (Ku)(x,t)  = g(x) + \int_0^t \int_I |W(x,y)| (u(y,s)-u(x,s)) \dy \, \mathrm ds\\
+ 2 \int_0^t \int_I W^-(x,y) u(y,s) \dy \, \mathrm ds    .
\end{multline*}
Consider $M_g$   the space of functions $u \in C([0,\tau];L^{\infty}(I))$ such that $u(x,0) = g(x)$, with the norm 
$\|v \|_{M_g} = \max_{t \in [0, \tau]}  \esssup_{x \in I} |v(x,t)| $.
Clearly $K: M_g \to M_g$. The goal is to show that $K$ is a contraction, for some sufficiently small $\tau$.
From the definition of $K$, 
\begin{multline*}
 (Ku - Kv)(x,t) =  \int_0^t \int_I |W(x,y)| (u(y,s)-v(y,s))  \dy \, \mathrm ds\\
 \qquad\qquad\qquad+ \int_0^t \int_I |W(x,y)| (-u(x,s)+v(x,s)) \dy \, \mathrm ds\\
+ 2 \int_0^t \int_I W^-(x,y) (u(y,s)-v(y,s)) \dy \, \mathrm ds    .
\end{multline*}
from which it is easy to see that 
\begin{multline*}
 |(Ku - Kv)(x,t) | \leq
  \| W \|_{\infty} \Big [ 3 \int_0^t \int_I | u(y,s)-v(y,s)|  \dy \, \mathrm ds \\
 +   \int_0^t |u(x,s)-v(x,s)| \, \mathrm ds \Big ] .
\end{multline*} 
This bound is the same as in the proof of \cite[Theorem~3.2]{medvedev2014nonlinear}, except for the factor 3 in the first term in the right-hand side. The conclusion then follows in a similar way: 
\begin{multline*} 3 \int_0^t \int_I | u(y,s)-v(y,s)|  \dy \, \mathrm ds 
 +   \int_0^t |u(x,s)-v(x,s)| \, \mathrm ds\\
 \leq 4\,  t \max_{s\in [0,t]} \| u(\cdot,s)-v(\cdot,s)\|_{L^\infty (I)} \, ,
 \end{multline*} 
 which does not depend on $x$,
 and hence
\begin{multline*}  \|  Ku - Kv \|_{M_g}  \leq 
4 \| W \|_{\infty} \max_{t\in[0,\tau]} t \max_{s\in [0,t]} \| u(\cdot,s)-v(\cdot,s)\|_{L^\infty (I)} \\
\leq  4 \| W \|_{\infty} \tau \max_{t\in[0,\tau]} \| u(\cdot,t)-v(\cdot,t)\|_{L^\infty (I)} 
=  4 \| W \|_{\infty} \tau \| u - v \|_{M_g} .
\end{multline*} 
Choosing $\tau = 1 / (8 \| W \|_{\infty})$, this ensures that $K$ is a contraction.
By Banach contraction theorem, this ensures existence and uniqueness of solution $u \in M_g$. The argument can then be iterated on further intervals $[k \tau, (k+1) \tau]$ to extend the result to the real line.
Furthermore, since the integrand (in time) in the definition of $K$ is continuous, $u$ is continuously differentiable. Thus, we have a classical solution. 
\end{IEEEproof}

\section{Bounds on the approximation errors}\label{sect:convergence-solutions}
The objective of this section is to compare the solutions of \eqref{DynGraphRep} to the solutions of \eqref{DynGraphonRep} (and analogously for solutions of \eqref{DynGraphOpp} and of \eqref{DynGraphonOpp}) for large values of $n$. Since these solutions belong to different spaces, it is necessary to explain how this comparison will be made. The following lemma is instrumental to a fair comparison.

\begin{lem}[Graph dynamics as graphon dynamics] \label{lem:equivalenza-discreto}
Let $A^{(n)}$ be the adjacency matrix of a graph, and $W_n(x,y)=\sum_{i,j} A^{(n)}_{ij}\mathds{1}_{I_i}(x)\mathds{1}_{I_j}(y)$ be its associated graphon,
where $I_i=(\frac{i-1}{n},\frac{i}{n}]$, $\mathds{1}_{I_i}(x)=1$ if $x\in I_i$ and 0 otherwise.  
Define 
$g_n(x) = \sum_{i=1}^n  g_i^{(n)} \mathds{1}_{I_i}(x)$,
and
$u_n(x,t)=\sum_{i=1}^n u^{(n)}_i(t)\mathds{1}_{I_i}(x)$.

If  $u^{(n)}(t)$ is solution of \eqref{DynGraphRep}, then $u_n(x,t)$ is  solution of 
\begin{equation}  \label{DynGraphRep-integrodiff}
\begin{cases}\displaystyle
    \frac{\partial u_n}{\partial t}(x,t)=n\alpha_n\int_{I}W_n(x,y)(u_n(y,t)-u_n(x,t))\dy  \\
    u_n(x,0)=g_n(x). 
\end{cases}
\end{equation}
Analogously, if $u^{(n)}(t)$ is solution of \eqref{DynGraphOpp}, then $u_n(x,t)$ is solution of 
\begin{equation}  \label{DynGraphOpp-integrodiff}
\begin{cases}\displaystyle
    \frac{\partial u_n}{\partial t}(x,t)=n\alpha_n\Big(\hspace{-1pt}\int_{I}W_n(x,y)u_n(y,t)\dy   -\hspace{-3pt} \int_{I}\! |W_n(x,y)|u_n(x,t)\dy\!\Big)  \\
    u_n(x,0)=g_n(x).
\end{cases}
\end{equation}
\end{lem}
\begin{IEEEproof}
We detail the proof for the repelling model.
From the definition of $u_n(x,t)$, we have
$$
\frac{\partial u_n(x,t)}{\partial t}=\sum_{i=1}^n 
\frac{\mathrm du^{(n)}_i(t)}{\mathrm dt} \, \mathds{1}_{I_i}(x),
$$
which means that, for any $i$, for all $x\in I_i$,
\begin{align*}
    \frac{\partial u_n(x,t)}{\partial t}=\frac{\mathrm du^{(n)}_i(t)}{\mathrm dt}&=
    \sum_{j=1}^n \alpha_n \, A_{ij}(u^{(n)}_j(t)-u^{(n)}_i(t))\\
    &= \sum_{j=1}^n n\alpha_n\hspace{-2pt}\int_{I_j}\hspace{-2pt}W_n(x,y)(u_n(y,t)-u_n(x,t))\dy \\
    &= n\alpha_n\int_{I} W_n(x,y)(u_n(y,t)-u_n(x,t))\dy. 
\end{align*}
The proof for the opposing model is analogous. 
\end{IEEEproof}
Notice that \eqref{DynGraphRep-integrodiff} and \eqref{DynGraphOpp-integrodiff} are particular cases of \eqref{DynGraphonRep} and \eqref{DynGraphonOpp}, respectively, where the graphon $W$ and the initial condition $g$ are piece-wise constant. Hence, Theorem~\ref{thm:existence-and-unicity} on existence and uniqueness of classical solutions applies to \eqref{DynGraphRep-integrodiff} and \eqref{DynGraphOpp-integrodiff} as well.
This implies that \eqref{DynGraphRep} and \eqref{DynGraphOpp} respectively have the equivalent formulations \eqref{DynGraphRep-integrodiff} and \eqref{DynGraphOpp-integrodiff} which use graphons, so that we can now directly compare $u(x,t)$ and $u_n(x,t)$. 

Before stating the main theorem (Theorem~\ref{thm:bound}), we prove a property that will be useful in the theorem's proof.
\begin{prop}\label{prop:quad_form_bound}
    Let $W(x,y):I^2\to [-1,1]$ be a signed graphon and $f\in L^2(I)$ be a function. Then, it holds
    \begin{equation*}
        \Big|\int_{I^2}(W(x,y)f(x)^2-W(x,y)f(x)f(y))\dx\dy\Big|
        \leq 2\|d_{|W|}\|_{\infty}\|f\|^2_2.
\end{equation*}%\hfill$\square$
\end{prop}
\begin{IEEEproof}By the triangle inequality,
    \begin{align}\label{passaggio_lemma}
        &\Big|\int_{I^2}(W(x,y)f(x)^2-W(x,y)f(x)f(y))\dx\dy\Big|\nonumber\\
        &\leq \int_{I^2}|W(x,y)f(x)^2|\dx\dy+\int_{I^2}|W(x,y)f(x)f(y)|\dx\dy.
    \end{align}
For the first term, using Tonelli's Theorem and Hölder's inequality, we can write
\begin{multline*}\int_{I^2}|W(x,y)f(x)^2|\dx\dy 
=\int_{I}f(x)^2\big(\int_I|W(x,y)|\dy\big)\dx\\
=\int_{I}f(x)^2d_{|W|}(x)\dx
\leq \|d_{|W|}\|_{\infty}\|f^2\|_1=\|d_{|W|}\|_{\infty}\|f\|_2^2.
\end{multline*}

Using this expression and Hölder's inequality,  we can write the second term of \eqref{passaggio_lemma} as
\begin{align}        \label{passaggi_Wf(x)f(y)}&\int_{I^2}|W(x,y)| \, |f(x)| \, |f(y)|\dx\dy\\&=\int_{I^2} \left( \sqrt{|W(x,y)|}\, |f(x)| \right)  \left(\sqrt{|W(y,x)|} \, |f(y)|\right)\dx\dy\nonumber\\
    &\leq \|f\sqrt{|W|}\|_2^2=\int_{I^2}|W(x,y)f(x)^2|\dx\dy\leq\|d_{|W|}\|_{\infty}\|f\|_2^2.\nonumber
    \end{align}
Adding the two terms yields the statement.
\end{IEEEproof}

\begin{rem}[Spectrum of the graphon Laplacian operator]
For a non-negative graphon $W$,     Proposition~\ref{prop:quad_form_bound} implies  
$$\langle f, L_Wf\rangle\leq 2 \, \|d_{W}\|_{\infty} \, \|f\|^2_2 , $$
where $L_W:L^2(I^2) \to L^2(I^2)$ is the graphon Laplacian operator defined by $L_W f (x)=\int_IW(x,y)(f(x)-f(y))\dy$ and $\langle \cdot,\cdot\rangle$ denotes the $L^2$ inner product.
$L_W$ is a self-adjoint operator with domain $L^2(I)$ 
and hence its  spectrum $\sigma(L_W)$  satisfies
$\inf \sigma(L_W) = \inf_{f: \|f\|_2=1} \langle f, L_W f \rangle$ and
$\sup \sigma(L_W) = \sup_{f: \|f\|_2=1} \langle f, L_W f \rangle$
(see e.g.~\cite[Theorem~2.19]{teschl-book}).
Hence, by Proposition~\ref{prop:quad_form_bound}, together with the fact that $\langle f, L_W f \rangle \ge  0$ for all $f$ (see \cite{bonnet2022consensus}),
$$\sigma(L_W) \subseteq [0, 2 \, \|d_{W}\|_{\infty}].$$
This is the extension to graphons of the well-known result for Laplacian matrices of graphs, whose eigenvalues lie in the interval $[0, 2 \, d_{(n)}]$ where $d_{(n)}$ is the maximum degree (as it is easy to show by Gershgorin Circle Theorem). \hfill $\square$
\end{rem}

We are now ready to state the main result of this section.
We will use the notation  $ \| (u-u_n)(\cdot,t)\|_2$ for the $L^2$ norm 
of $u-u_n$ in the $x$ variable only.

\begin{thm}[Approximation error estimates]\label{thm:bound}
Let $W: I^2 \to [-1,1]$ be a signed graphon, and let 
$g \in L^\infty(I)$, 
$W_n$ and $g_n$ be as in Lemma~\ref{lem:equivalenza-discreto}.
If $u_n$ and $u$ are solutions of  \eqref{DynGraphRep-integrodiff} and \eqref{DynGraphonRep}, respectively,  then for all $t \in [0,T]$
\begin{multline*}  
    \| (u-u_n)(\cdot,t)\|_2^2
    \\ \leq\hspace{-3pt} \left(\hspace{-2pt}  \|g-g_n\|_2+ \frac{C_{u,T}}{n\alpha_n\,\|d_{|W_n|}\|_{\infty}} \vertiii{T_{W-n\alpha_nW_n}}\hspace{-2pt} \right) \hspace{-2pt}\exp(2n\alpha_n\,\|d_{|W_n|}\|_{\infty}T),
\end{multline*}
where $C_{u,T} = \displaystyle \esssup_{t\in[0,T], x\in I} |u(x,t)|$ (which is finite by Theorem~\ref{thm:existence-and-unicity}).

If $u_n$ and $u$ are solutions of  \eqref{DynGraphOpp-integrodiff} and \eqref{DynGraphonOpp}, respectively,  then for all $t \in [0,T]$
\begin{multline*}    \| (u-u_n)(\cdot,t)\|_2^2 \leq 
     \exp(4 n\|d_{|W_n|}\|_{\infty}\alpha_n T)
     \Big [
     \|g-g_n\|_2   \\ +
         \frac{C_{u,T} }{n\alpha_n\|d_{|W_n|}\|_{\infty}} \! \left( \vertiii{T_{W^+} -n\alpha_nT_{W_n^+}} \!+\! \vertiii{T_{W^-} -n\alpha_nT_{W_n^-}}  \right) \Big],
\end{multline*}
where $W = W^+ - W^-$ denotes the decomposition of $W$ in its positive part $W^+ = \max(W,0)$ and its negative part $W^- = \max(-W, 0)$.%\hfill$\square$
\end{thm}
\begin{IEEEproof} 
    We will use the short-hand notations 
    $\xi_n(x,t)=u_n(x,t)-u(x,t)$
    and
    $\tilde W_n(x,y) = n\alpha_nW_n(x,y) - W(x,y)$. We begin by the repelling case.
From \eqref{DynGraphonRep} and \eqref{DynGraphRep-integrodiff}, we have 
    \begin{multline*}
        \frac{\partial \xi_n(x,t)}{\partial t}
        = n\alpha_n\int_I W_n(x,y)(u_n(y,t)-u_n(x,t)) \dy\\
        - \int_I W(x,y) (u(y,t)-u(x,t)) \dy.
    \end{multline*}
We add  the following zero term to the right-hand side:
$-n\alpha_n\int_I W_n(x,y) (u(y,t)-u(x,t)) \dy 
+ n\alpha_n\int_I W_n(x,y) (u(y,t)-u(x,t)) \dy $, 
to obtain
    \begin{multline*}
        \frac{\partial \xi_n(x,t)}{\partial t}
        = n\alpha_n\int_I W_n(x,y)(\xi_n(y,t)-\xi_n(x,t)) \dy\\
        +\int_I \tilde W_n(x,y) (u(y,t)-u(x,t)) \dy.
    \end{multline*}
Then, we multiply  
both sides by $\xi_n(x,t)$ and integrate over $I$ 
\begin{multline}    \label{eq:proof-convergence-int}
\int_I \frac{\partial \xi_n(x,t)}{\partial t} \,\xi_n(x,t) \dx =\\
n\alpha_n\int_{I^2} W_n(x,y)(\xi_n(y,t)-\xi_n(x,t))\xi_n(x,t) \dx \dy\\
    +\int_{I^2} \tilde W_n(x,y)(u(y,t)-u(x,t))\xi_n(x,t) \dx\dy.
\end{multline}
For the left-hand side, notice that
\begin{equation} \label{eq:proof-convergence-int-left}
\int_I \frac{\partial \xi_n(x,t)}{\partial t} \xi_n(x,t) \dx 
= \frac{1}{2} \int_I \frac{\partial (\xi_n(x,t))^2}{\partial t}  \dx
= \frac{1}{2} \frac{\mathrm d}{\mathrm d t} \| \xi_n(\cdot,t)\|_2^2.
\end{equation}

By using Proposition~\ref{prop:quad_form_bound} with $f(x)=\xi(x,t)$, 
we estimate the first term of the right-hand side of~\eqref{eq:proof-convergence-int} as follows:
\begin{multline} \label{eq:proof-convergence-int-right1} \Big| n\alpha_n\int_{I^2} W_n(x,y)(\xi_n(y,t)-\xi_n(x,t))\xi_n(x,t) \dx \dy \Big|    
 \\
\leq 2n\alpha_n\|d_{|W_n|}\|_{\infty}\|\xi_n\|_2^2,
\end{multline}
while for the second term we have
\begin{multline}  \label{eq:proof-convergence-right2-preliminary}
  \Big| \int_{I^2} \tilde W_n(x,y)(u(y,t)-u(x,t))\xi_n(x,t) \dx\dy \Big| \\
\leq 
  \Big| \int_{I^2} \tilde W_n(x,y) u(y,t) \xi_n(x,t) \dx\dy \Big| \\
  + \Big| \int_{I^2} \tilde W_n(x,y) u(x,t) \xi_n(x,t) \dx\dy \Big| .
\end{multline}
Recalling the definition of the integral operator $T_W$ associated with a graphon $W$, we have
\begin{multline*}
     \Big| \int_{I^2} \tilde W_n(x,y) u(y,t) \xi_n(x,t) \dx\dy \Big| \\=
\Big| \int_{I} \left( T_{\tilde W_n} \xi_n(\cdot,t) \right)(y) u(y,t)  \dy \Big| 
\leq \| (T_{\tilde W_n} \xi_n)(\cdot,t) \, u(\cdot,t)  \|_1 .
\end{multline*}
Then, by H\"older inequality,
\begin{multline*}
\| (T_{\tilde W_n} \xi_n)(\cdot,t) \, u(\cdot,t)  \|_1 
\leq \| (T_{\tilde W_n} \xi_n)(\cdot,t) \|_2  \, \| u(\cdot,t)  \|_2  \\
\leq \vertiii{T_{\tilde W_n}} \, \|\xi_n(\cdot,t) \|_2 \, \| u(\cdot, t) \|_2
\leq \vertiii{T_{\tilde W_n}} \, \|\xi_n(\cdot,t) \|_2 \, \| u(\cdot, t) \|_\infty.
\end{multline*}
Similarly, for the second term in \eqref{eq:proof-convergence-right2-preliminary},
\begin{multline*}
     \Big| \int_{I^2} \tilde W_n(x,y) u(x,t) \xi_n(x,t) \dx\dy \Big| \\=
\Big| \int_{I} \left( T_{\tilde W_n} (\xi_n(\cdot,t) u(\cdot,t)) \right)(y)  \dy \Big| 
\leq \|  T_{\tilde W_n} (\xi_n(\cdot,t) u(\cdot,t))  \|_1 ,
\end{multline*}
and
\begin{multline*}
\| T_{\tilde W_n} (\xi_n(\cdot,t) u(\cdot,t)) \|_1 
\leq \| T_{\tilde W_n} (\xi_n(\cdot,t) u(\cdot,t)) \|_2  \\
\leq \vertiii{T_{\tilde W_n}} \, \|\xi_n(\cdot,t) u(\cdot,t)\|_2 
\leq \vertiii{T_{\tilde W_n}} \, \|\xi_n(\cdot,t) \|_2 \, \| u(\cdot, t) \|_\infty.
\end{multline*}
We have obtained that for the second term in \eqref{eq:proof-convergence-int}:
\begin{multline}
    \label{eq:proof-convergence-right2}
  \Big| \int_{I^2} \tilde W_n(x,y)(u(y,t)-u(x,t))\xi_n(x,t) \dx\dy \Big| \\
  \leq
2 \vertiii{T_{\tilde W_n}} \, \|\xi_n(\cdot,t) \|_2 \, \| u(\cdot, t) \|_\infty
  \leq 2 \, C_{u,T} \,  \vertiii{T_{\tilde W_n}} \, \|\xi_n(\cdot,t) \|_2 ,
\end{multline}
with $C_{u,T} = \esssup_{t \in [0,T], x\in I} |u(x,t) | $,
which is finite, since $u$ is continuous in time and essentially bounded in space, by Theorem~\ref{thm:existence-and-unicity}.

Combining \eqref{eq:proof-convergence-int}, 
\eqref{eq:proof-convergence-int-left}, \eqref{eq:proof-convergence-int-right1} and \eqref{eq:proof-convergence-right2}, we have
\begin{multline}  \label{eq:proof-convergence-final}
   \frac{1}{2} \frac{\mathrm d}{\mathrm d t} \| \xi_n(\cdot,t)\|_2^2
   \\\leq 2 \,n\alpha_n\,\|d_{|W_n|}\|_{\infty}   \| \xi_n(\cdot,t)\|_2^2
   + 2 \, C_{u,T} \, \vertiii{T_{\tilde W_n}} \, \|\xi_n(\cdot,t) \|_2 .
\end{multline}

This is  similar to the estimate obtained in \cite[Eq.~(4.16)]{medvedev2014nonlinear}, but with different constants, $\vertiii{T_{\tilde W_n}}$ instead of $\|\tilde W_n\|_2$, and the presence of the scaling factor $\alpha_n$.
Then, the argument can then be concluded along the same lines as \cite{medvedev2014nonlinear}, as follows. 
To `divide by $\| \xi_n(\cdot, t) \|_2$' avoiding the risk to divide by zero, one
defines an auxiliary function 
$\phi_{\delta}(t)=\sqrt{\| \xi_n(\cdot, t) \|_2^2+\delta}$, for any $\delta>0$.
From \eqref{eq:proof-convergence-final},
\[
     \frac{1}{2} \frac{\mathrm d}{\mathrm dt}\phi_{\delta}(t)^2 \leq 
     2 n\alpha_n\,\|d_{|W_n|}\|_{\infty}\phi_{\delta}(t)^2 + 2 C_{u,T} \, \vertiii{T_{\tilde W_n}} \phi_{\delta}(t)
\]
and, by chain rule and dividing by $\phi_{\delta}(t)$, 
\[
\frac{\mathrm d}{\mathrm dt}\phi_{\delta}(t)
\leq  2n\alpha_n \phi_{\delta}(t) + 2 C_{u,T}\, \vertiii{T_{\tilde W_n}} .
\]
%------------------------------
With notation $a = 2n\alpha_n\,\|d_{|W_n|}\|_{\infty}$, $b = 2 C_{u,T}\, \vertiii{T_{\tilde W_n}}$,
we have
\[
\frac{\mathrm d}{\mathrm dt}\phi_{\delta}(t)
\leq  a \phi_{\delta}(t) + b.
\] 
By Grönwall's lemma, $\phi_{\delta}(t) \leq \zeta(t)$
where $
\frac{\mathrm d}{\mathrm dt}\zeta(t)
=  a \zeta(t) + b$
and $\zeta(0) = \phi_{\delta}(0)$.
Solving the ODE, we have
\[ \zeta(t) = 
e^{a t} \left( \phi_{\delta}(0) + \tfrac{b}{a} (1 - e^{-a t}) \right) \,.\]
Since $a>0$, $b>0$, $\phi_{\delta}(0)>0$ and $t \in [0,T] $,
\begin{multline*}
    \zeta(t) \leq  
e^{a T} \left( \phi_{\delta}(0) + \tfrac{b}{a} \right) 
\\= e^{2n\alpha_n\,\|d_{|W_n|}\|_{\infty}T} \left( \phi_{\delta}(0) +  \frac{C_{u,T}}{n\alpha_n\,\|d_{|W_n|}\|_{\infty}}\, \vertiii{T_{\tilde W_n}} \right) \,.
\end{multline*} 
Since $\phi_{\delta}(t) \leq \zeta(t)$ and $\delta$ is arbitrary, this gives the conclusion.
%--------------------------------

\bigskip
Now we consider the opposing case. 
From \eqref{DynGraphonOpp} and \eqref{DynGraphOpp-integrodiff}, 
recalling that $|W_n(x,y)| = W_n(x,y) + 2 W_n^-(x,y)$,
we have 
    \begin{multline*}
        \frac{\partial \xi_n(x,t)}{\partial t}
        = n\alpha_n\int_I |W_n(x,y)| \, (u_n(y,t)-u_n(x,t)) \dy\\
        - \int_I |W(x,y)| \, (u(y,t)-u(x,t)) \dy \\
        - 2n\alpha_n \int_I W_n^-(x,y) u_n(y,t) \dy
        + 2 \int_I W^-(x,y) u(y,t) \dy.
    \end{multline*}
We add  the following zero term to the right-hand side:
$-n\alpha_n\int_I {|}W_n(x,y){|} (u(y,t)-u(x,t)) \dy 
+ n\alpha_n\hspace{-1pt}\int_I \hspace{-1pt}{|}\hspace{-1pt}W_n(x,y)\hspace{-1pt}{|} (u(y,t)\hspace{-1pt}-\hspace{-1pt}u(x,t)) \dy 
- 2 n\alpha_n\hspace{-1pt}\int_I\hspace{-1pt} W_n^-(x,y) u(y,t) \dy 
+ 2 n\alpha_n\int_I W_n^-(x,y) u(y,t) \dy$. 
Then we multiply both sides by $\xi_n(x,t)$ and integrate in $I$ in the $x$ variable. %\margin{togliere da qui in poi?}
We obtain:
\begin{align}
\label{eq:proof-convergence-int_Opp}
\nonumber  \int_I &\frac{\partial \xi_n(x,t)}{\partial t} \,\xi_n(x,t) \dx\\
\nonumber=&n\alpha_n\int_{I^2} |W_n(x,y)| \, (\xi_n(y,t)-\xi_n(x,t))\xi_n(x,t) \dx \dy\\
\nonumber    &+\!\!\int_{I^2} (n\alpha_n|W_n(x,y)| - |W(x,y)| )(u(y,t)-u(x,t))\xi_n(x,t) \dx\dy\\
\nonumber    &{-} 2n\alpha_n \int_{I^2} W_n^-(x,y) \xi_n(y,t) \xi_n(x,t) \dx\dy\\
    &{-} 2 \int_{I^2} ( n\alpha_nW_n^-(x,y) - W^-(x,y) )  u(y,t)  \xi_n(x,t) \dx\dy.
\end{align}

Compare \eqref{eq:proof-convergence-int_Opp} with  \eqref{eq:proof-convergence-int}.
The left-hand side is the same 
and the first two terms of the right-hand side are almost the same, with the only difference that $W_n(x,y)$ and $W(x,y)$ 
are now replaced by $|W_n(x,y)|$ and $|W(x,y)|$. 
Hence, \eqref{eq:proof-convergence-int-left}, \eqref{eq:proof-convergence-int-right1}
and \eqref{eq:proof-convergence-right2} still apply, except that 
$\vertiii{T_{\tilde W_n}}$ is replaced by 
$\vertiii{T_{n\alpha_n|W_n| - |W|}}$.
In the following, we take care of the remaining terms, which were not present in \eqref{eq:proof-convergence-int}.

Recalling the calculations done in $\eqref{passaggi_Wf(x)f(y)}$, we have
\begin{align*}
   &\Big| 2n\alpha_n\int_{I^2} W_n^-(x,y) \xi_n(y,t) \xi_n(x,t) \dx\dy \Big| \\
    &\leq 2 n\alpha_n\int_{I^2} | W_n^-(x,y)| | \xi_n(y,t) | |\xi_n(x,t) | \dx\dy\\
    &= 2 n\alpha_n\|d_{W_n^-}\|_{\infty} \|\xi_n(\cdot, t) \|_2^2\leq 2 n\alpha_n\|d_{|W_n|}\|_{\infty} \|\xi_n(\cdot, t) \|_2^2.
\end{align*}
For the last term,
\begin{align*}
   &\Big| 2 \int_{I^2}  ( n\alpha_nW_n^-(x,y) - W^-(x,y) )  u(y,t)  \xi_n(x,t) \dx\dy \Big| \\
   &\leq 2 \int_I |(T_{n\alpha_nW_n^- - W^-} \xi_n)(\cdot,t))(y)| \, |u(y,t)| \dy \\
   &= 2 \|(T_{n\alpha_nW_n^- - W^-} \xi_n)(\cdot,t) \, u(\cdot,t)\|_1\\
   &\leq 2 \|(T_{n\alpha_nW_n^- - W^-} \xi_n)(\cdot,t) \|_2  \, \|  u(\cdot,t)\|_{\infty} \\
   &\leq 2 \vertiii{T_{n\alpha_nW_n^- - W^-}} \, \|\xi_n(\cdot,t) \|_2  \, C_{u,T}.
\end{align*}
Collecting the upper bounds for all four terms in the right-hand side of \eqref{eq:proof-convergence-int_Opp}, we get
\begin{multline*}
\frac{1}{2} \frac{\mathrm d}{\mathrm d t} \| \xi_n(\cdot,t)\|_2^2 
   \,\leq\,  4 \,n\alpha_n\,\|d_{|W_n|}\|_{\infty}  \| \xi_n(\cdot,t)\|_2^2 \\
   + 2 \, C_{u,T} \, \left( \vertiii{T_{n\alpha_n|W_n| - |W|}} + \vertiii{T_{n\alpha_nW_n^- - W^-}} \right)\, \|\xi_n(\cdot,t) \|_2.
\end{multline*}
Using $|W| = W^+  +  W^-$, 
%$T_{W_n+W} = T_{W_n} + T_W$ 
$T_{W_1+W_2} = T_{W_1} + T_{W_2}$ 
and triangle inequality, we have
$\vertiii{T_{n\alpha_n|W_n|-|W|}} \leq \vertiii{T_{n\alpha_nW_n^+} -T_{W^+}} + \vertiii{T_{n\alpha_nW_n^-} -T_{W^-}} $. Hence,
\begin{multline*}
\frac{1}{2} \frac{\mathrm d}{\mathrm d t} \| \xi_n(\cdot,t)\|_2^2 
   \,\leq\,  4 \, n\alpha_n\,\|d_{|W_n|}\|_{\infty} \| \xi_n(\cdot,t)\|_2^2 \\
   + 4 \, C_{u,T} \, \left( \vertiii{T_{n\alpha_nW_n^+} -T_{W^+}} + \vertiii{T_{n\alpha_nW_n^-} -T_{W^-}} \right)\, \|\xi_n(\cdot,t) \|_2,
\end{multline*}
which is the analogous of \eqref{eq:proof-convergence-final}. 
The conclusion then follows by the same technique involving $\phi_{\delta}$ and Grönwall's lemma, with $a=4 \, n\alpha_n\,\|d_{|W_n|}\|_{\infty}$ and $b= 4 \, C_{u,T} \, \left( \vertiii{T_{n\alpha_nW_n^+} -T_{W^+}} + \vertiii{T_{n\alpha_nW_n^-} -T_{W^-}} \right)$. 
\end{IEEEproof}

Theorem~\ref{thm:bound} implies that, so long as 
the initial conditions converge and the finite graphs (weighted by $n\alpha_n$) converge to the graphon and have bounded degrees, then  the dynamics on the finite graphs converge to the dynamics on the graphon, in the sense that solutions converge on bounded intervals.

\section{Convergence of solutions for sampled graphs}
\label{sect:convergence-sampled}

The goal of this section is to apply Theorem~\ref{thm:bound} and provide conditions for solutions of opinion dynamics on large graphs that are sampled from graphons to converge,  as $n\to\infty$, to solutions of opinion dynamics on graphons. 
Subsection~\ref{subsect:conv-sampled-positive} is dedicated to the analysis of the term $\vertiii{T_{n\alpha_nW_n}-T_W}$ when $W$ is a non-negative graphon, proving its convergence to zero when $n\alpha_n=\eps_n^{-1}$ in Theorem~\ref{convergence_sampled_to_graphon_NEW}. In Subsection~\ref{subsect:bounded-degree}, we prove Theorem~\ref{thm:degrees}, stating that, for a non-negative graphon and for $n\alpha_n=\eps_n^{-1}$, the term $n\alpha_n\|d_{W_n}\|_\infty$ is bounded both from above and from below. %, as the bound in Theorem~\ref{thm:bound} could explode if that was not the case. 
Next, in Subsection~\ref{subsect:extension-to-signed}, we extend these two theorems to signed graphons. As a last step, in Subsection~\ref{subsect:initial-conditions}, we consider the term $\|g-g_n\|_2$ relative to the initial conditions, and show its convergence to zero in Proposition~\ref{prop:conv_in_cond}. In Subsection~\ref{subsect:final-thm}, we combine the previous results to derive sufficient conditions for convergence in Theorem~\ref{thm:sampled-convergence}.
These conditions include choosing $\eps_n=\omega(\log n/n).$ Here, $\omega$ and the symbols $o,\; O,\; \Omega$ and $\Theta$, used throughout this section, are the Bachmann–Landau notations.\footnote{The Bachmann–Landau symbols are defined as follows: $f(n)=o(g(n))$ if $\frac{f(n)}{g(n)}\to 0$ for $n\to\infty$; $f(n)=\omega(g(n))$ if
$g(n)=o(f(n))$; $f(n)=O(g(n))$ if there exist constants $M,N$ s.t.\ $|f(n)|\leq M|g(n)|$ for $n>N$; $f(n)=\Omega(g(n))$ if $g(n)=O(f(n))$; and $f(n)=\Theta(g(n))$ if both $f(n)=O(g(n))$ and $f(n)=\Omega(g(n)).$}

\subsection{Convergence of sampled graphs -- non-negative graphon}\label{subsect:conv-sampled-positive}
In this subsection, we study the convergence to zero  of $\vertiii{T_{\eps_n^{-1}W_n}-T_W}$. Almost sure convergence of sampled graphs to graphons is well-known under very mild assumptions, but most results are given in `cut distance', i.e., in cut norm but up to a suitable unknown measure-preserving transformation (see e.g.~\cite{lovasz2012book}, Coroll.~11.34 for the case with deterministic latent variables and Prop.~11.32 for the case with stochastic latent variables). This  is equivalent to almost sure convergence in cut norm, up to an unknown permutation of vertices of each graph in the sequence of graphs \cite[Lemma 5.3]{borgs2008}.
Having chosen points $X_i$ in an increasing order, it is natural to hope that vertices are already ordered in the most useful way and might not need further permutation to obtain convergence. 
Moreover, the equivalence between operator norm and cut norm from~\cite[Lemma E.6]{janson2013graphons} does not apply because $\eps_n^{-1}W_n$ can be unbounded. 
The desired convergence in operator norm is proved in \cite{avella2018centrality} under some regularity assumption on the graphon. \begin{prop}[\!\!{\cite[Thm.~1]{avella2018centrality}}]
\label{convergence_sampled_to_graphon_AvellaMedina}
    Let $W:I^2\to[0,1]$ be a piece-wise Lipschitz graphon and $W_n$ a graph sampled from $W$ with either deterministic or stochastic latent variables. 
    If $\eps_n = n^{-\tau}$ for some $\tau \in [0,1)$,
    then almost surely $\vertiii{T_{\eps_n^{-1}W_n}-T_W}\to0$ for $n\to\infty$.\hfill$\square$
\end{prop}

In this subsection, we will obtain Theorem~\ref{convergence_sampled_to_graphon_NEW}, which shows that the same holds true under minimal assumptions. To do this, we build upon several results from the literature, going beyond the most classical proof techniques as in \cite{lovasz2012book}, which involve subgraph densities and only give results in cut distance.

Before starting, we observe that, when $W$ is a graphon taking values in $[0,1]$, so that $(W_n - W)$ takes values in $[-1,1]$, convergence in $L^1$ norm implies convergence in operator norm, thanks to the following result. 
\begin{prop}
    \label{prop:convergence_barW_W_op}Let   $ W:I^2 \to [-1,1]$ be a signed graphon. Then, it holds that $\vertiii{T_W}\leq\sqrt{\| W\|_1}$.
\end{prop}

\begin{IEEEproof}
    Let $W$ be a signed graphon, we have
    $$\vertiii{T_{ W}}^2=\sup_{\|u\|_2=1}\|T_{ W}u\|_2^2=\sup_{\|u\|_2=1}\int_I\left(\int_I W(x,y)u(y)\dy\right)^2\dx.$$
    Using the Cauchy-Schwarz inequality and $\|u\|_2=1$, we obtain 
    \begin{align*}   \left(\int_I W(x,y)u(y)\dy\right)^2\leq\| W(x,\cdot)\|_2^2 \,\|u\|_2^2=
    \int_I W(x,y)^2\dy. 
    \end{align*}
   Since $W$ takes values in $[-1,1]$, it holds that $W(x,y)^2\leq|W(x,y)|$, from which 
$$\vertiii{T_{W}}^2\leq\int_{I\times I} W(x,y)^2\dx\dy \leq\int_{I\times I}| W(x,y)|\dx\dy=\| W\|_1.$$    
\end{IEEEproof}
Having chosen deterministic or stochastic latent variables $X_1,\dots, X_n$ as in Definition~\ref{latent_var}, we will consider the weighted graph $\bar G_n$ having $n$ vertices and adjacency matrix $\bar A^{(n)}$ with $\bar A^{(n)}_{ij} = W(X_i,X_j)$. 
We will denote by $\bar W_n$ the associated piece-wise constant graphon 
$\bar W_n(x,y) = \sum_{i,j} W(X_i,X_j) \mathds{1}_{I_i}(x)\mathds{1}_{I_j}(y)$. To prove the convergence of $\eps_n^{-1}W_n$ to $W$, we will use $\bar W_n$ as an intermediary.

We start by studying the convergence of $\bar W_n$ to $W$.
This is done in a different way (and under different assumptions) for the case of deterministic and of stochastic latent variables.

In the case with deterministic latent variables, the following convergence result holds.
\begin{prop} 
\label{prop:convergence_barW_W_det}
Let $W:I^2 \to [0,1]$ be an almost everywhere continuous graphon and let  $\bar W_n(x,y) = \sum_{i,j} W(X_i,X_j) \mathds{1}_{I_i}(x)\mathds{1}_{I_j}(y)$, where $X_i = i/n$. Then, 
\[ \| \bar W_n - W \|_1 \to 0 \text{ for $n \to \infty$.} \]
\end{prop}
\begin{IEEEproof}
The proof uses the following simple argument, mentioned in \cite{medvedev2014nonlinear}, from the proof of \cite[Lemma 2.5]{borgs2011}. 
The first step is to notice that for  almost every point $(x,y)$ there is
pointwise convergence of $\bar W_n(x,y)$ to $W(x,y)$.
Indeed, for almost every $(x,y)$, $W$ is continuous at $(x,y)$,
and $\bar W_n(x,y) = W(x_n,y_n)$ for some $(x_n,y_n)$ such that $\| (x_n,y_n) - (x,y) \|_2 \leq \sqrt 2 / n \to 0$ (since $(x_n,y_n) = (i/n, j/n)$ where $i,j$ are the indexes such that $(x,y) \in I_i \times I_j$), and hence $\bar W_n(x,y) \to W(x,y)$.
From this pointwise (almost everywhere) convergence, the desired $L^1$ convergence is obtained by Lesbesgue dominated convergence theorem, having noticed that $|\bar W_n(x,y) - W(x,y)| \leq 2$.
\end{IEEEproof}

In the case with stochastic latent variables, convergence of $\bar W_n$ to $W$ is obtained in \cite[Theorem 2.14(a)]{borgs2019Lp}, where it is stated for a wider class of graphons, possibly unbounded, which includes as a particular case the more classical graphons considered in this paper. Notice that, with stochastic latent variables, no regularity assumption is needed. 
\begin{prop}[{\!\!\cite[Theorem 2.14(a)]{borgs2019Lp}}]
\label{prop:convergence_barW_W_stoc}
Let  $W:I^2 \to [0,1]$ be a graphon 
and let $\bar W_n(x,y) = \sum_{i,j} W(X_i,X_j) \mathds{1}_{I_i}(x)\mathds{1}_{I_j}(y)$ where $X_1, \dots, X_n$ are the stochastic latent variables. 
Then
\( \| \bar W_n - W \|_1 \to 0 \text{ a.s., for $n \to \infty$.} \)
\hfill$\square$
\end{prop}

After having studied the convergence of $\bar W_n$ to $W$, one needs to study the convergence of $\eps_n^{-1} W_n$ to $\bar W_n$. 
Notice that $\eps_n \bar A^{(n)}$ is the expectation of $ A^{(n)}$ conditioned on $X_1, \dots, X_n$. Hence, convergence of $\eps_n^{-1} A^{(n)}$ to $\bar A^{(n)}$ is  a concentration result that has been obtained in the literature about inhomogeneous Erd\H{o}s-R\'enyi graphs.
\begin{prop}[\!\!{\cite[Theorem~5.2]{lei2015}}]\label{prop:lei}
    Let $A^{(n)}$ be a random adjacency matrix with expected value $\eps_n \bar A^{(n)}$ and independently drawn edges. 
    Let $b_n=n\eps_n\max_{ij}\bar A^{(n)}_{ij}$. If $b_n=\Omega(\log n)$, then, for any $r>0$, there exists a constant $C>0$ such that $$\Pr(\vertiii{A^{(n)}-\eps_n\bar A^{(n)}}\leq C\sqrt{b_n})\geq 1-n^{-r},$$ where $\vertiii{A^{(n)}}$ is the spectral norm of the matrix $A^{(n)}$.~$\hfill\square$
\end{prop}

\begin{rem}   
The spectral norm of an adjacency matrix is related to the operator norm of the associated graphon through the relationship $\vertiii{T_{W_n}}=\frac{1}{n}\vertiii{A^{(n)}}$.  
Indeed, since $A^{(n)}$ is self-adjoint, $\vertiii{A^{(n)}}$ is equal to the greatest  eigenvalue (in absolute value) of $A^{(n)}$. 
Similarly, $\vertiii{T_{W_n}}$ is equal to the greatest  eigenvalue (in absolute value) of $T_{W_n}$, see e.g.\ \cite{gao2019spectral}. Finally, it is easy to see that $\lambda$ is an eigenvalue of $A^{(n)}$ if and only if $\lambda/n$ is an eigenvalue of $T_{W_n}$.~$\hfill\square$ 
\end{rem}

By noticing that $\max_{ij}\bar A^{(n)}_{ij}=\|\bar W_n\|_\infty$, we can use Proposition~\ref{prop:lei} to obtain the following result, where we denote $b_n=n\eps_n\max_{ij}\bar A^{(n)}_{ij}=n\eps_n\|\bar W_n\|_\infty$.
\begin{prop}
  \label{prop:operator-norm-concentration}
    Let $W:I^2 \to [0,1]$ be a graphon and let $\bar W_n$ and $W_n$ be the above-described weighted graph and sampled graph, with either deterministic or stochastic latent variables.
    Then, for $n\to \infty$, if $b_n=\Omega(\log n)$, then \vspace{-2mm}
$$ \vertiii{T_{\eps_n^{-1}W_n}-T_{\bar W_n}}\to 0\quad \text{a.s.} $$
\end{prop}

\begin{IEEEproof}
    We have \begin{equation}\label{norm-operator-to-matrix}
    \vertiii{T_{\eps_n^{-1}W_n}-T_{\bar W_n}}=\frac{1}{n\eps_n}\vertiii{A^{(n)}-\eps_n\bar A^{(n)}}.\end{equation} 
From Proposition~\ref{prop:lei}, taking $r>1$ we can use Borel-Cantelli Lemma to get $\vertiii{A^{(n)}-\eps_n\bar A^{(n)}}=O(\sqrt{b_n})$ almost surely.
 Since $b_n=n\eps_n\|{\bar W}_n\|_\infty\leq n\eps_n$, substituting this inequality in \eqref{norm-operator-to-matrix} yields \begin{equation}\label{eq:dependence-of-convergence-on-eps}
     \vertiii{T_{\eps_n^{-1}W_n}-T_{\bar W_n}}=O\Big(\frac{\sqrt{b_n}}{n\eps_n}\Big)=O\Big(\frac{1}{\sqrt{n\eps_n}}\Big)\to0\quad\text{a.s.}
\end{equation}
\end{IEEEproof}

Notice that Proposition~\ref{prop:operator-norm-concentration} requires to assume $b_n = \Omega(\log n)$ to obtain convergence of $\eps_n^{-1} W_n$ to $\bar W_n$.
The following remark will show that such assumption is often satisfied. Moreover, Proposition~\ref{prop:conv-for-W-null} will allow us to study the case in which this assumption does not hold;  in this case, we will directly obtain the convergence of $\eps_n^{-1} W_n$ to $W$, without using $\bar W_n$ as an intermediary.

\begin{rem}
\label{rem:condition-for-convergence} 
Provided $\eps_n=\omega(\log n/n)$, the condition ``$b_n=\Omega(\log n)$" in Proposition~\ref{prop:operator-norm-concentration} is always satisfied when $\|W\|_1> 0$. Indeed, $\|\bar W_n\|_\infty\geq\|\bar W_n\|_1\to\|W\|_1$ by Propositions~\ref{prop:convergence_barW_W_det} and~\ref{prop:convergence_barW_W_stoc} and the reverse triangle inequality. 
Therefore, $b_n=n\eps_n\|\bar W_n\|_\infty=\omega(\log n)\Omega(1)=\omega(\log n)$. $\hfill\square$
\end{rem}

\begin{prop}
   \label{prop:conv-for-W-null}Let $W:I^2\to[0,1]$ 
    be a graphon and $W_n$ be a graph sampled from $W$, with either deterministic (under the additional assumption that $W$ is a.e.\ continuous) or stochastic latent variables. 
    If $\eps_n=\omega(\log n/n)$ and  $b_n=O(\log n)$,
    then almost surely $\vertiii{T_{\eps_n^{-1}W_n}-T_W}\to0$ for $n\to\infty$.
\end{prop}

\begin{IEEEproof}
By Remark~\ref{rem:condition-for-convergence}, if $b_n=O(\log n/n)$, then necessarily $\|W\|_1=0$, which in turn implies $\vertiii{T_W}=0$ by Proposition~\ref{prop:convergence_barW_W_op}. 
Next, define $d_{(n)}$ and $\bar d_{(n)}$ as the maximum degree of, respectively, $A^{(n)}$ and $\bar A^{(n)}$. 
Notice that $\frac{1}{n}\bar d_{(n)}\leq \max_{ij}\bar A^{(n)}_{ij}$, and that, since $b_n=O(\log n)$, we have  $\max_{ij}\bar A^{(n)}_{ij}=O\big(\frac{\log n}{n\eps_n}\big)\to0$.  As we will show later in Lemma~\ref{lemma3improved},  $\frac{1}{n\eps_n}|d_{(n)}-\eps_n\bar d_{(n)}|\to 0$ a.s., which, since $\frac{1}{n}\bar d_{(n)}\to0$, implies $\frac{1}{n\eps_n}d_{(n)}\to0$ a.s..  
Furthermore, by Gershgorin's Circle Theorem, $\vertiii{T_{\eps_n^{-1}W_n}}=\frac{1}{n\eps_n}\vertiii{A^{(n)}}\leq\frac{1}{n\eps_n}d_{(n)}$, which goes to 0. The convergence $\vertiii{T_{\eps_n^{-1}W_n}}\to0$, together with $\vertiii{T_W}=0$, yields the statement. 
\end{IEEEproof}

We now have all the ingredients to obtain 
the following convergence result, analogous to Proposition~\ref{convergence_sampled_to_graphon_AvellaMedina} under milder assumptions.
\begin{thm}[Convergence of sampled graphs] \label{convergence_sampled_to_graphon_NEW}
Let $W:I^2\to[0,1]$ 
    be a graphon and $W_n$ be a graph sampled from $W$, with either deterministic (under the additional assumption that $W$ is a.e.\ continuous) or stochastic latent variables.  
    If 
    $\eps_n=\omega(\log n/n)$, 
    then almost surely $\vertiii{T_{\eps_n^{-1}W_n}-T_W}\to0$ for $n\to\infty$.
\end{thm}  

\begin{IEEEproof}
If ${b_n} =\Omega(\log n)$, then Propositions~\ref{prop:convergence_barW_W_det}, \ref{prop:convergence_barW_W_stoc} and~\ref{prop:convergence_barW_W_op} combined together give $\vertiii{\bar W_n-W}\to0$ a.s., which, when further combined with 
Proposition~\ref{prop:operator-norm-concentration}, yields the result by the triangle inequality. 
If $b_n=O(\log n)$, the result is given by Proposition~\ref{prop:conv-for-W-null}. 
Finally, if $b_n$ is neither $\Omega(\log n)$, nor $O(\log n)$, then 
fix any $c_0>0$ and consider the two subsequences, one with the terms $\geq c_o \log n$ and the other with the terms $< c_0 \log n$.  
Applying the arguments above to these subsequences yields the result.
\end{IEEEproof}

\subsection{Bounds on the degree of sampled graphs}\label{subsect:bounded-degree}
In the bounds of Theorem~\ref{thm:bound}, the term $n\alpha_n\|d_{|W_n|}\|_\infty$ appears both as  exponent and as denominator. 
%In this subsection, 
We will show that if $n\alpha_n=\epsilon_n^{-1}$ and $\eps_n = \omega(\log n / n)$, then $n\alpha_n\|d_{|W_n|}\|_\infty$ is almost surely  bounded both from above and from below, which implies that, for $n$ that goes to infinity, $(n\alpha_n\|d_{|W_n|}\|_\infty)^{-1}$ and $\exp(n\alpha_n\|d_{|W_n|}\|_\infty T)$ are almost surely bounded (for all finite $T$ in the latter expression). These bounds are based on the following analysis of the distribution of the degrees of sampled graphs, which we detail in this subsection for the case of non-negative graphons  (Theorem~\ref{thm:degrees}) and then, in the next subsection, extend to signed graphons. 

We start with the bound from above. Consider the graph $G_n$, sampled from the non-negative graphon $W$, and the associated piece-wise constant graphon $W_n$. 
We consider the nodes' degrees $d_i$, and use the label $(i)$ to denote the degrees reordered in increasing order  $d_{(1)}\leq\dots\leq d_{(n)}$. 
 We define the normalized degrees $\delta_{i}=\frac{d_{i}}{n}$ and
 notice that $\delta_{(n)}$ is equal to $\|d_{W_n}\|_\infty$. Indeed, due to $W_n$ being piece-wise constant, we have $$\|d_{W_n}\|_\infty=\esssup_{x\in I}d_{W_n}(x)=\max_{x\in I}d_{W_n}(x)=\max_i\frac{d_i}{n}=\delta_{(n)}.$$
Similarly, $\|d_{\bar W_n}\|_\infty  =\bar \delta_{(n)}$, where $\bar\delta_{(i)}=\frac{\bar d_{(i)}}{n}$, with  $\bar d_{(1)}\hspace{-2pt}\leq\dots\leq \bar d_{(n)}$ being the reordered degrees of $\bar G_n$.

In order to study $\delta_{(n)}$, we need the following lemma. 
\begin{lem}
\label{lemma3improved}
Let $W:I^2\to[0,1]$ be a graphon. If $n$ is large enough, then with probability at least $1-\nu$ the normalized degrees  of graphs $\bar G_n$ and $G_n$ sampled from $W$  
satisfy:
\begin{equation}\label{gamma_new}
\max_{i=1,\dots, n} \vert \eps_n^{-1} \delta_{(i)}-  \bar \delta_{(i)}\vert \leq 
\sqrt{\dfrac{ \log(2n/\nu)}{n \, \eps_n}}.
\end{equation}
\end{lem}

\begin{IEEEproof}
This is \cite[Lemma~I]{garin2024corrections}, extended by the straightforward addition of the scaling  factors $\eps_n$.
\end{IEEEproof}

With this, we are ready to prove that $\eps_n^{-1}\|d_{W_n}\|_\infty$ is bounded from above. 
\begin{prop}\label{prop:degree-upper-bound}
Let $W:I^2\to[0,1]$ be a graphon and $W_n$ be a graph sampled from $W$, with either  deterministic or stochastic latent variables. If $\eps_n = \omega(\log n / n)$, then $$\epsilon_n^{-1}\|d_{W_n}\|_\infty=O(1) \, \text{ a.s.}$$
\end{prop}
\begin{IEEEproof}
From Lemma~\ref{lemma3improved}, 
if we take $\nu = n^{-a}$ for some $a>1$ and use Borel-Cantelli Lemma, 
we obtain almost sure convergence: 
\begin{equation}\label{lem:Lemma-concentration-BorelCantelli} \max_i |\eps_n^{-1} \delta_{(i)}-\bar \delta_{(i)}| = O\left( \sqrt{\frac{\log n}{n \, \eps_n}}\right)  \; \text{a.s.}. \end{equation}
    Taking $\eps_n = \omega(\log n / n)$, $\max_i |\eps_n^{-1} \delta_{(i)}-\bar \delta_{(i)}| \to 0$ almost surely. Since this holds for all $i$,  it holds in particular for $i=n$, which we have shown to be $\delta_{(n)}=\|d_{W_n}\|_\infty$. We have $$ \Big|\eps_n^{-1} \|d_{W_n}\|_\infty- \|d_{\bar W_n}\|_\infty\Big| \to 0 \text{ a.s.}.$$
Since $\|d_{\bar W_n}\|_\infty\leq1$, this gives  $\eps_n^{-1} \|d_{W_n}\|_\infty=O(1)$ a.s., concluding the proof.
\end{IEEEproof}

\smallskip
Moving on to the lower bound, we start by pointing out that the average (normalized) degree of a piece-wise constant graphon is equal to its $L^1$ norm. Indeed,
\begin{align*}
\|W_n\|_1&=\int_{I^2}|W_n(x,y)|\dx\dy
    =\sum_{i,j}\int_{I_i\times I_j}|W_n(x,y)|\dx\dy\\
    &=\sum_{i,j}\frac{1}{n^2}|W(X_i,X_j)| =\frac{1}{n}\sum_i \sum_j\frac{W(X_i,X_j)}{n}=\frac{1}{n}\sum_i \frac{d_i}{n}.
\end{align*}
As the maximum degree is greater than or equal to the average degree, we have $$\delta_{(n)}=\|d_{W_n}\|_\infty\geq\|W_n\|_1.$$
We now show that $\eps^{-1}_n\|W_n\|_1\to \|W\|_1$.  
For this, we have convergence of $\|\bar W_n\|_1$ to $\|W\|_1$ (Propositions~\ref{prop:convergence_barW_W_det} and~\ref{prop:convergence_barW_W_stoc}), and we now introduce a lemma that states the convergence of $\eps_n^{-1}\| W_n\|_1$ to $\|\bar W_n\|_1$. It is a concentration result, obtained by recalling that  $\bar W_n$ is the expectation of $W_n$ (conditioned on $X_1,\ \dots,\ X_n$ in the case of stochastic latent variables).

\begin{lem} \label{prop:concentration-norm1W}
Let $W:[0,1]^2\to[0,1]$ be a graphon and $W_n$ be a graph sampled from $W$, with either stochastic or deterministic latent variables. Then, for any $t\in(0,1)$
$$\Pr (|\eps_n^{-1}\|W_n\|_1-  \|\bar W_n\|_1|\geq t \|\bar W_n\|_1)\leq 2e^{-t^2\eps_n \frac{n^2}{6} \|\bar W_n\|_1}.$$
If $\|\bar W_n\|_1$ is not vanishing for $n\to\infty$, and if $n\eps_n\to\infty$, then
\[  \left|\eps_n^{-1} \|W_n\|_1 - \|\bar W_n\|_1 \right| \to 0\; \text{a.s.}\]
\end{lem}
\begin{IEEEproof}
All along this proof, in the case of stochastic latent variables, probabilities and expectations are meant conditioned on $X_1, \dots, X_n$.

Recall that
    $\frac{n^2}{2} \|W_n\|_1 = E_n = \sum_i \sum_{j>i} \eta_{ij}$, where $\eta_{ij}$ are independent Bernoulli random variables, $\eta_{ij}$ having mean $\eps_n W(X_i,X_j)$. Notice that there are $m $ terms in the summation, where $m = \frac{n(n-1)}{2} \leq \frac{n^2}{2}$.
Then notice  that $ \eps_n \frac{n^2}{2} \|\bar W_n\|_1 = 
\sum_i \sum_{j>i} \eps_n W(X_i,X_j) = \E(E_n)$.
By the Chernoff bound for independent Bernoulli random variables, we have, for any $t\in(0,1)$
\begin{align*}
    \Pr (|E_n-\E E_n|\geq t \E E_n)&\leq 2e^{-t^2\E E_n/3},\\
    \Pr (|\tfrac{n^2}{2} \|W_n\|_1-\eps_n \tfrac{n^2}{2} \|\bar W_n\|_1|\geq t \eps_n \tfrac{n^2}{2} \|\bar W_n\|_1)&\leq 2e^{-t^2\eps_n \frac{n^2}{2} \|\bar W_n\|_1/3},\\
     \Pr (|\eps_n^{-1}\|W_n\|_1-  \|\bar W_n\|_1|\geq t \|\bar W_n\|_1)&\leq 2e^{-t^2\eps_n \frac{n^2}{6} \|\bar W_n\|_1}.
\end{align*}

We want both $t \|\bar W_n\|_1$ and $2\exp(-t^2\eps_n \frac{n^2}{6} \|\bar W_n\|_1)$ to go to zero. 
To achieve this, we can choose $t$ such that $t^2\|\bar W_n\|_1^2=\frac{1}{n\eps_n}$, which,  if $\|\bar W_n\|_1$ is not vanishing, implies that $t\in(0,1)$ for a sufficiently large $n$. 
Indeed, since $\|\bar W_n\|_1\leq 1$, then $t^2\|\bar W_n\|_1\geq t^2\|\bar W_n\|_1^2$, thus, $-t^2\eps_n \frac{n^2}{6} \|\bar W_n\|_1\leq -t^2\eps_n \frac{n^2}{6} \|\bar W_n\|_1^2=-\frac{n}{6}$, which makes the exponential go to zero summably. 
Since with this choice of $t$ we also have that $t\|\bar W_n\|_1=\frac{1}{\sqrt{n\eps_n}}\to0$, we can apply Borel-Cantelli Lemma to obtain the statement.
\end{IEEEproof}

 With this lemma, we can obtain the result we need.

\begin{prop}\label{prop:degree-lower-bound}
    Let $W:I^2\to[0,1]$ be a graphon, such that $\|W\|_1>0$, and $W_n$ be a graph sampled from $W$, with either deterministic (under the additional assumption that $W$ is a.e.\ continuous) or stochastic latent variables. 
    Then $\eps_n^{-1}\|W_n\|_1=\Omega(1)$ almost surely.
\end{prop}
\begin{IEEEproof}
Throughout this proof, convergences are meant almost surely.
Propositions~\ref{prop:convergence_barW_W_det}
and~\ref{prop:convergence_barW_W_stoc} imply $\|\bar W_n-W\|_1\to 0$ and thus, by reverse triangle inequality, $| \| \bar W_n \|_1 - \|  W \|_1 |\to0$, which implies $\| \bar W_n \|_1\to\|  W \|_1$. 
If $\|W\|_1>0$, then, from Lemma~\ref{prop:concentration-norm1W}, 
    $| \eps_n^{-1}\|W_n\|_1-\|\bar W_n\|_1 | \to 0$. 
Combining the two convergences, we have $\eps_n^{-1}\|W_n\|_1\to\| W\|_1>0$. As $\|W\|_1$ is a constant, this yields the statement.
\end{IEEEproof}

Since the maximum degree is greater than the average degree, $\eps_n^{-1}\|W\|_1\leq\eps_n^{-1}\|d_{W_n}\|_\infty$, and combining  Proposition \ref{prop:degree-upper-bound} and Proposition \ref{prop:degree-lower-bound}, we obtain the main result of this section.
\begin{thm}\label{thm:degrees}
Let $W:[0,1]^2\to[0,1]$ be a graphon, such that $\|W\|_1>0$, and $W_n$ be a graph sampled from $W$, with either deterministic (under the additional assumption that $W$ is a.e.\ continuous) or stochastic latent variables. If $\eps_n=\omega(\log n/n)$,  
    then almost surely  $\eps_n^{-1}\|d_{W_n}\|_\infty=\Theta(1)$ and $\eps_n^{-1}\|W\|_1=\Theta(1)$, where $\|d_{W_n}\|_\infty$ and $\|W\|_1$ are respectively the maximum and the average normalized degrees of the sampled graph.\hfill$\square$
\end{thm}

\subsection{Signed graphons: sampled graphs and degree}\label{subsect:extension-to-signed}
In the previous Subsections~\ref{subsect:conv-sampled-positive} and~\ref{subsect:bounded-degree}, we have shown, for non-negative graphons, that the sequence of sampled graphs converges to the graphon and that the degree is bounded. In this subsection, we show that these results also hold for signed graphons. 

To extend the results of Subsection~\ref{subsect:conv-sampled-positive}, we highlight the equivalence between the positive and negative parts of a graph sampled from a signed graphon, and the graphs sampled from the positive and negative parts of a signed graphon.

\begin{rem}
    Let $W=W^+-W^-$ be a signed graphon and $A^{(n)}=A^+-A^-$ the adjacency matrix of a signed graph sampled from $W$, where the superscripts ``+'' and ``-'' refer to the positive and negative parts of $W$ and $A^{(n)}$, respectively. %Let $A^+$ and $A^-$ be the positive and negative parts of $A^{(n)}$, such that $A^{(n)}=A^+-A^-$.
    Let $\tilde{A}_+$ and $\tilde{A}_-$ be the adjacency matrices of graphs sampled from, respectively, $W^+$ and $W^-$. Then, it is clear that $\tilde{A}_+$, $\tilde{A}_-$, and
    $\tilde{A}^{(n)}=\tilde{A}_+-\tilde{A}_-$  have the same distributions of 
$A^+$, $A^-$, and $A^{(n)}$, respectively.    \hfill$\square$
\end{rem}

This remark, together with Theorem~\ref{convergence_sampled_to_graphon_NEW}, implies the following proposition.
\begin{prop}\label{prop:convergence-sampled-from-signed}
    Let $W:I^2\to[-1,1]$ 
    be a signed graphon and $W_n$ a graph sampled from $W$, with either deterministic (under the additional assumption that $W$ is a.e.\ continuous) or stochastic latent variables.  
    If 
    $n\eps_n \to \infty$,
    then almost surely $\vertiii{T_{\eps_n^{-1}W^+_n}\hspace{-2pt}-T_{W^+}}\to0$, $\vertiii{T_{\eps_n^{-1}W^-_n}-T_{W^-}}\to0$, and $\vertiii{T_{\eps_n^{-1}W_n}-T_W}\to0$ for $n\to\infty$.~\hfill$\square$
\end{prop}

Moving on to the extension of the results of Subsection~\ref{subsect:bounded-degree}, all the results concern the norm of the degree $\|d_{W_n}\|_\infty$, where $W_n$ is sampled from a non-negative graphon $W$. In the bounds of Theorem~\ref{thm:bound}, it is actually the term $\|d_{|W_n|}\|_\infty$ that appears, where $W_n$ is sampled from a signed graphon. 
\begin{rem}
Let $W$ be a signed graphon, $A^{(n)}$ be the adjacency matrix of a graph sampled from $W$, and $\tilde{\tilde{A}}^{(n)}$ be the adjacency matrix of a graph sampled from $|W|$. Then, $\tilde{\tilde{A}}^{(n)}$ and $|A^{(n)}|$ (the absolute value being applied entry-wise) have the same distribution.\hfill$\square$
\end{rem}
This remark and Theorem~\ref{thm:degrees} imply the following result.
\begin{prop}\label{prop:degrees-signed}
    Let $W:I^2\to[-1,1]$ be a signed graphon such that $\|W\|_1>0$, and $W_n$ be a graph sampled from $W$, with either deterministic (under the additional assumption that $W$ is a.e.\ continuous) or stochastic latent variables. If $\eps_n=\omega(\log n/n)$,  
    then $\eps_n^{-1}\|d_{|W_n|}\|_\infty=\Theta(1)$ and $\eps_n^{-1}\|W\|_1=\Theta(1)$  almost surely.\hfill$\square$
\end{prop}

\subsection{Convergence of initial conditions}\label{subsect:initial-conditions}
In order to have convergence of the solutions of the two different dynamics, we need to have compatible initial conditions. 
Given the initial condition $g(x)$ for the dynamics on the graphon, we choose to define the initial condition of the dynamics on the sampled graph $g_n(x)$ as a piece-wise constant function by assigning the value $g(X_i)$ to  the entire interval $I_i$. We now show that with this definition, the term $\|g_n-g\|_2$ present in Theorem \ref{thm:bound} goes to zero as $n$ increases.
\begin{prop}[Convergence of initial conditions]\label{prop:conv_in_cond}
    Given $g: I \to \R$ 
    and given $X_1, \dots, X_n \in I$, define the piece-wise constant function
    $g_n(x)=g(X_i)$ for all $x\in I_i$.
\begin{enumerate}
    \item If $X_1, \dots, X_n$ are deterministic latent variables, 
if $g$ is bounded and almost everywhere continuous,
 then $\lim_{n \to \infty} \|g_n-g\|_2 =0$.
 \item If $X_1, \dots, X_n$ are stochastic latent variables, if $g \in L^2(I)$, then
$\lim_{n \to \infty} \|g_n-g\|_2 =0$ almost surely.%\hfill$\square$
 \end{enumerate}
\end{prop}

 \begin{IEEEproof}
 1) Preliminarily, observe that since $g$ is almost everywhere continuous, for almost all $x$ we have
 $\lim_{h \to 0} |g(x+h)-g(x)| = 0$.
In the case of deterministic latent variables, for all $i$, for all $x \in I_i$, $g_n(x) = g(X_i)$, with $|X_i - x| \le \frac{1}{n}$, which goes to zero when $n \to \infty$. 
Hence, for almost all $x$ we have  $\lim_{n \to \infty} |g_n(x)-g(x)| = 0$.
From this pointwise (almost everywhere) convergence we can obtain convergence in $L^2$ norm, using the Lebesgue dominated convergence theorem.
\smallskip

2) The proof in the case of stochastic latent variables is inspired by the proof of  \cite[Theorem 2.14(a)]{borgs2019Lp}, but with significant modifications, to obtain convergence in $L^2$ norm instead of $L^1$ norm (different choice of the function used to approximate $g$ and different proof of its convergence).

\smallskip
A first part of the proof is to define a suitable function $\gamma_m$ and prove that $\|\gamma_m - g \|_2 \to 0$ for $m \to \infty$.
Consider the positive and negative parts $g^+$ and $g^-$ of $g$, and define $\gamma_m(x) = \gamma_m^+(x) - \gamma_m^-(x)$, 
where 
\[ \gamma^+_m(x) = 
    \sum_{j=1}^m \bar \gamma^+_j \, \ind_{J_j}(x), \qquad
    \bar \gamma^+_j = \sqrt{m \, \int_{J_j} (g^+(\xi))^2 \dxi } 
    \]
is a piece-wise constant function over intervals
$J_1 = [0, \frac{1}{m}]$ and $J_j = (\frac{j-1}{m} , \frac{j}{m}]$ for $j=2,\dots, m$,
and $\gamma_m^-$ has an analogous definition, just replacing $g^+$ with $g^-$.
The notation $\gamma_m = \gamma_m^+ - \gamma_m^-$ is adopted for convenience, but clearly $\gamma_m^+$ and $\gamma_m^-$ might not be the positive and negative parts of $\gamma_m$.

The assumption $g \in L^2(I)$ also implies that $g^+ \in L^2(I)$ 
and hence $(g^+)^2$ is integrable.
Since
$(\gamma_m^+(x))^2 = m \int_{J_j} (g^+(\xi))^2 \dxi $ with $J_j$ an interval of length $1/m$ containing $x$, and $(g^+)^2$ is integrable, then when $m\to\infty$ by Lebesgue differentiation theorem  $ (\gamma_m^+(x))^2\to (g^+(x))^2$ pointwise a.e.;
$g^+$ being non-negative, this further implies
$ \gamma_m^+(x)\to g^+(x)$ pointwise almost everywhere. 
Now notice that $\|\gamma_m^+\|_2^2 = 
\sum_j \frac{1}{m} (\bar \gamma_j^+)^2 =
\sum_j \int_{J_j} (g^+(\xi))^2 \dxi = \|g^+\|_2^2.$
This, together with pointwise a.e. convergence, ensures convergence in $L^2$ norm (i.e., $\|\gamma_m^+ - g^+\|_2\to0$) by an extension of Scheffé's Lemma, which is proved as follows. 
The function defined by
$\psi_m(x) = 4\,[(\gamma_m^+(x))^2 + (g^+(x))^2] -
    (\gamma_m^+(x) - g^+(x))^2$ is non-negative,
since
$(\gamma_m^+(x) - g^+(x))^2 \leq (2\, \max (\gamma_m^+(x), g^+(x)))^2 = 4 \max((\gamma_m^+(x))^2, (g^+(x))^2)
\leq 4 \, [(\gamma_m^+(x))^2 + (g^+(x))^2]$.
By applying Fatou's Lemma to $\psi_m$, we have
$\int_I \liminf_m \psi_m(x) \dx \leq \liminf_m \int_I \psi_m(x) \dx$.
Thanks to pointwise convergence a.e., the left-hand side is equal to $4 \|g^+\|_2^2 + 4 \|g^+\|_2^2 -0$.
Thanks to $\|\gamma_m^+\|_2 =\|g^+\|_2 $ for all $m$, the right-hand side is equal to
$4 \|g^+\|_2^2 + 4 \|g^+\|_2^2 - \limsup_m \|\gamma_m^+ - g^+\|_2^2$.
Hence, we have obtained $\limsup_m \|\gamma_m^+ - g^+\|_2^2 \leq 0$ and, since clearly 
$\liminf_m \|\gamma_m^+ - g^+\|_2^2 \geq 0$, we conclude that 
$\lim_m \|\gamma_m^+ - g^+\|_2^2 = 0$.
The same proof steps can be used to show that $\lim_m \|\gamma_m^- - g^-\|_2 = 0$ and hence also
$\lim_m \|\gamma_m - g\|_2 = 0$ by triangle inequality.

\smallskip The final goal of this proof is to show that  $\lim_{n \to \infty} \|g_n-g\|_2 =0$ a.s.,
which can be equivalently rephrased as: for all $\eps > 0$,
$\Pr \left( \limsup_{n}{\|g_n-g\|_2} \leq \eps \right) = 1$.

Fix an arbitrary $\eps>0$, and fix $m$ such that
$\|\gamma_m - g\|_2< \eps/2$ 
(which surely exists, since we have shown above that $\|\gamma_m - g\|_2 \to 0$ for $m \to \infty$). For the rest of the proof $m$ will be fixed, and $\gamma_m$ will be denoted $\gamma$ so as to emphasize this.

For any $n$, recall the notation $I_1, \dots, I_n$: $I_i = (\frac{i-1}{n},\frac{i}{n}]$, 
and define $\tilde g_n(x) = \sum_{i=1}^n \gamma(X_i) \ind_{I_i}(x)$.
The next step is to show that a.s.\ $\| \tilde g_n - \gamma \|_1 \to 0$ for $n \to \infty$.
To do so, define $N_j$  the number of points from $X_1, \dots, X_n$ that fall within the interval $J_j$. 
Notice that $N_j \sim \text{Binomial}(n,\frac{1}{m})$ and hence it has expected value $\frac{n}{m}$ and, by Hoeffding inequality, for any $t_n>0$,
$ \Pr(|N_j - n/m| \geq t_n) \leq 2 \exp(-2 t_n^2 / n) $
and hence 
$ \Pr(\bigcup_{j=1,\dots,m} \{|N_j - n/m| \geq t_n\}) \leq 2 m \exp(-2 t_n^2 / n) $. By taking 
$t_n$ such that $t_n = o(n)$ and $\exp(-2 t_n^2 / n)$ goes to zero fast enough to be summable
(for example $t_n = n^{2/3}$), we can apply Borel-Cantelli Lemma     and obtain that, for $n\to\infty$, almost surely for all $j$ we have $N_j = \frac{n}{m} + o(n)$. 
Now recall that $\tilde g_n(x) $ is constant equal to 
$ \bar \gamma_1 $ on intervals $I_1,\dots, I_{N_1}$,  
equal to $\bar \gamma_2$ on intervals $I_{N_1+1},\dots, I_{N_1+N_2}$, \dots, equal to 
$\bar \gamma_j$ on intervals $I_{N_{1}+\dots+N_{j-1}+1},\dots, I_{N_{1}+\dots+N_{j-1}+N_{j}}$,
 having taken the notation $\bar \gamma_j = \bar \gamma_j^+ - \bar \gamma_j^-$.
Compare this with $\gamma(x)$, which is constant equal to 
$ \bar \gamma_1 $ on intervals 
$I_1,\dots, I_{\lfloor n/m\rfloor}$,
 equal to $\bar \gamma_2$ on intervals 
$I_{\lceil n/m\rceil + 1}, \dots, I_{\lfloor 2n/m\rfloor}$, \dots, equal to     $\bar \gamma_j$ on intervals $I_{\lceil(j-1) n/m\rceil + 1}, \dots, I_{\lfloor j \, n/m \rfloor}$.
Notice that in case $m/n$ is not integer there are $m$ intervals $I_{\lceil j n/m \rceil}$ ($j=1,\dots,m$) that fall across  $J_j$ and $J_{j+1}$ and where $\gamma$ might take two different values.
Counting these $m$ intervals and all intervals
where $\gamma(x)$ and $\tilde g(x)$ take a different constant value, we find at most 
$m \, |N_1-\lceil n/m \rceil| +  
(m-1) \, |N_2- 2 \lceil n/m \rceil|
+ \dots +
 |N_m- m \lceil n/m \rceil|  $,
which almost surely is $o(n)$ for $n \to \infty$.   
Hence, almost surely the number of intervals (each of length $\frac{1}{n}$) where $\gamma - \tilde g_n$ is not ensured  to be zero is $o(n)$.
Finally,
$\gamma$ has a finite maximum  $\Gamma$ (since it is piece-wise constant and each value $\bar \gamma_j$ is finite because $g^+, g^- \in L^2(I)$),
and hence on the intervals where it is non-zero, $|\tilde g (x) - \gamma(x)| \leq 2 \Gamma$.
In conclusion, almost surely for $n\to \infty$ we have $\|\gamma - \tilde g_n\|_1 \leq \frac{1}{n} 2 \Gamma o(n) = o(1)$.

Finally, we consider  
$\|g_n-\tilde g_n\|_2^2 
= \int_I \sum_{i=1}^n (g(X_i)-\gamma(X_i))^2  \ind_{I_i}(x) \dx
= \frac{1}{n} \sum_{i=1}^n (g(X_i)-\gamma(X_i))^2$, which tends a.s. to its expectation $ \int_I (g(x)-\gamma(x))^2 \dx = \|g-\gamma\|_2^2$ 
by the Strong Law of Large Numbers.

The conclusion is obtained using the triangle inequality 
\[ \|g-g_n\|_2 \leq \|g-\gamma\|_2 + \|\gamma-\tilde g_n\|_2
        + \|\tilde g_n - g_n\|_2\]
and recalling the three  intermediate results obtained above:
$\|g-\gamma\|_2 \leq \eps/2$, 
and almost surely $\lim_{n \to \infty} \|\gamma-\tilde g_n\|_2 = 0$ and $\lim_{n \to \infty} \|\tilde g_n - g_n\|_2 = \|g-\gamma\|_2 \leq \eps/2$.
This implies that almost surely $\limsup_{n \to \infty} \|g-g_n\|_1 \leq \eps$ and
concludes the proof.
\end{IEEEproof}

Notice that the assumption that $g$ is bounded and a.e.\ continuous cannot be removed in the case of deterministic latent variables: indeed, these are the assumptions under which Riemann sums converge to the Riemann integral of~$g$. 
Random latent variables allow for lighter assumptions, 
in relation with the fact that Riemann sums using random intervals  converge a.s.\ to the integral under milder assumptions, given by the Strong Law of Large Numbers.

\subsection{Convergence of solutions}\label{subsect:final-thm}

We are now ready to combine the results from this section, namely Propositions~\ref{prop:convergence-sampled-from-signed}, \ref{prop:degrees-signed} and~\ref{prop:conv_in_cond}, with Theorem~\ref{thm:bound} to conclude our main convergence result.
\begin{thm}[Convergence of solutions on sampled graphs]\label{thm:sampled-convergence}
Let $W:I \to [-1,1]$ be a signed graphon such that $\|W\|_1>0$,  and let $u$ be a solution to \eqref{DynGraphonRep} (respectively, to \eqref{DynGraphonOpp}). For each $n\in\mathbb{N}$, 
consider stochastic latent variables $X_1, \dots X_n$ as in Definition~\ref{latent_var} and
let $W_n$ be a graph with $n$ nodes, sampled from $W$ as per Definition~\ref{sampledgraph_negativeweights}
with $\eps_n$ such that $\eps_n = \omega(\log n / n)$, 
and let $u_n$ be the solution to \eqref{DynGraphRep} (respectively, to \eqref{DynGraphOpp}). 
Assume that the initial conditions are $g\in L^\infty(I)$ for  \eqref{DynGraphonRep} and  \eqref{DynGraphonOpp},
and $g_n(x) = \sum_i g(X_i) \ind_{I_i}(x)$ for 
\eqref{DynGraphRep} and \eqref{DynGraphOpp}. 
If $\alpha_n = \frac{1}{n \eps_n}$, then for any fixed $T >0$ almost surely
\[ \max_{t \in [0,T]}  \|u_n(\cdot, t)-u(\cdot,t)\|_2\to 0 \quad  \text{for $n \to \infty$.} \]
If moreover $W$ is almost everywhere continuous and $g$ is almost everywhere continuous and bounded, then the same holds also in the case with deterministic latent variables. \hfill$\square$
\end{thm}
The correct choice of the scaling relationship between $\alpha_n$ and $\eps_n$ is crucial to prevent the right-hand sides of \eqref{DynGraphRep} and \eqref{DynGraphOpp} from either diverging or vanishing for large $n$, and therefore to obtain a meaningful convergence result (on bounded intervals of length $T$). Notice that $n\eps_n$ is almost surely the average degree of the sampled graph, per Proposition~\ref{prop:degrees-signed}. Hence $\alpha_n$ can be interpreted as ensuring that agents are influenced by averages of their neighbors opinions, as opposed to sums of the opinions (not weighted by the number of interactions). 

\section{Discussion}\label{sect:discussion}

We devote this section to a detailed comparison between our results and prior work, both by ourselves and by other authors. 

This paper 
significant extends our 
preliminary account~\cite{cdc2024}. Indeed, we include both a speed parameter $\alpha_n$ in the dynamics and a sampling parameter $\epsilon_n$ in sampling the graphs, which allow for much greater generality in the sparsity of the sampled graphs (in~\cite{cdc2024} we had $\epsilon_n=1$ and $\alpha_n=\frac1n$, whereas here $\eps_n=\omega(\log n/n)$ and $\alpha_n=\frac1{n\eps_n}$). In order to achieve this generality, 
a significantly revised analysis was necessary because a straightforward addition of the parameter $\alpha_n$ in the proof of \cite[Theorem~2]{cdc2024} would result in a term $n\alpha_n$ in the argument of the exponential, which would make the bound diverge when set equal to $\eps_n^{-1}$. Instead, the term $n\alpha_n\|d_{|W_n|}\|_{\infty}$ is constant when $n\alpha_n=\eps_n^{-1}$, as shown in Section~\ref{subsect:bounded-degree}, so that the bound can go to 0 when the conditions of Theorem~\ref{thm:sampled-convergence} are satisfied.
Moreover, in Section~\ref{sect:convergence-sampled} we make much weaker (in fact, the weakest possible) assumptions on the regularity of the graphon.

More generally, these same elements (presence of scaling parameters to accommodate sparse graphs and no regularity requirements) distinguish our work from the rest of the literature.
A key reference is the seminal work by Medvedev \cite{medvedev2014nonlinear}. 
Compared to the latter, we restrict ourselves to linear dynamics, but we consider signed graphons, we add the scalings $\alpha_n$ and $\eps_n$, 
and we obtain a stronger upper bound, which involves $\vertiii{T_{W-n\alpha_nW_n}}$, whereas \cite{medvedev2014nonlinear} considers $\| W-W_n \|_2$. Having the operator norm is essential to apply the result to sampled graphs, for which $\vertiii{T_{W-n\alpha_nW_n}}$ vanishes for large $n$ under mild assumptions, 
while this is not the case for $\| W-W_n \|_2$. 
Indeed, a counterexample is provided \cite[Ex.~2.3]{medvedev2014nonlinear}, which considers the graphs sampled from the constant graphon $W(x,y) \equiv p$ (i.e., the Erd\H{o}s-R\'enyi graphs). 

Multiple types of sampled graphs were considered in Medvedev's later paper \cite{medvedev2019}. 
Albeit with a different definition of the sampling probability (equal to $W(X_i,X_j)$ in our paper, and equal to $n^2\int_{I_i\times I_j}W(x,y)\dx\dy$ in \cite{medvedev2019}), our definition of sampled graphs is covered by the analysis of \cite{medvedev2019}, which however requires $W$ to be non-negative and more restrictive sampling weights $\eps_n=\omega(n^{-1/2})$. The latter limitation has recently been removed in \cite{nagpal2024synchronization}, where the authors allow for sampling weights as small as $\eps_n=\omega(\log n/n)$ and obtain the stronger $L^\infty$ convergence, but require the graphon to be non-negative and $C^1$. Moreover, \cite{nagpal2024synchronization} presents convergence results in probability, instead of almost surely as in this paper.

\section{Simulations}\label{sect:simulations}
\begin{figure}
    \centering
    \includegraphics[width=0.835\columnwidth]{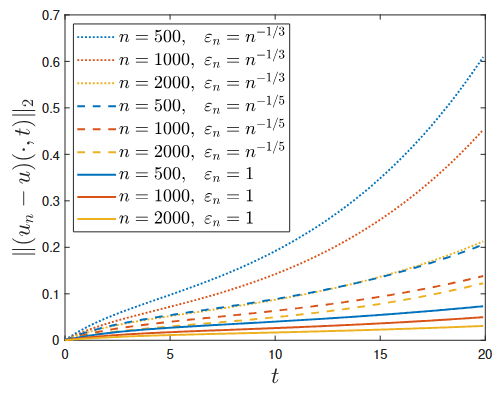}
    \includegraphics[width=0.835\columnwidth]{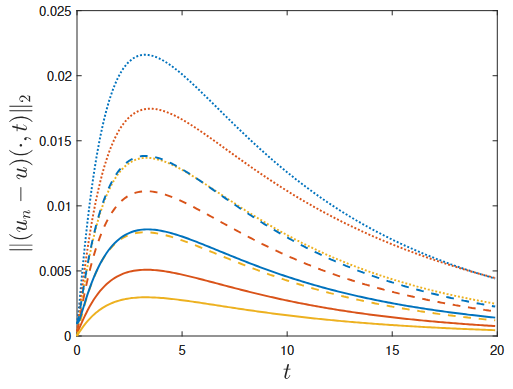}
    \caption{Evolution in time of the $L^2$ norm of the approximation error. Top: repelling dynamics. Bottom: opposing dynamics.}
    \label{fig:simulations}
\end{figure}

In this section,
we illustrate our results with simulations. 
We consider the same setting as in \cite{cdc2024}, but we add the parameter $\eps_n$ (which can be considered equal to 1 in \cite{cdc2024}) in the sampling procedure and $\alpha_n=(n\eps_n)^{-1}$ in the dynamics. In Figure~\ref{fig:simulations}, we show the plot of the approximation errors for different values of $n$ and different definitions of $\eps_n$. As expected from Theorem~\ref{thm:sampled-convergence}, the error decreases with $n$, but can increase with time. 
We can observe that the approximation errors for the repelling and the opposing dynamics show different behaviors. The former grows unbounded with time, whereas the latter grows for a while before decreasing to zero. This difference is explained by the fact that, in our examples, the solutions of the repelling dynamics are unbounded, thus allowing for unbounded errors, whereas the solutions of the opposing dynamics converge to zero, thus eventually making the error go to zero as well.  Illustrative plots of the mentioned solutions can be found in \cite{cdc2024} for $\eps_n=1$.
We also observe that the error is smaller when $\eps_n$ is larger. This is consistent with our analysis; indeed, inspecting the proofs of convergence reveals that the only dependence on $\eps_n$ appears in expression \eqref{eq:dependence-of-convergence-on-eps}, which gives a decay rate
$O\big(\frac{1}{\sqrt{n\eps_n}}\big)$.

\section{Conclusion}\label{sect:conclusion}

In this paper we considered two models of opinion dynamics with antagonistic interactions (the repelling and the opposing model), and we extended them to graphons. We showed existence and uniqueness of solutions and stated bounds on the error between graph  and graphon solutions. 
We then showed that these bounds go to zero if we take a sequence of (possibly sparse) graphs sampled from a graphon, as the number of nodes goes to infinity.

These results prove graphons to be a helpful tool to handle large networks  with antagonistic interactions, as we have shown how dynamics on the graphon well approximate their finite-dimensional versions. A line of future work, which we have started in~\cite{prisant:ecc2025}, is understanding the properties of the graphon dynamics itself, such as its behavior as time goes to infinity.

%\bibliographystyle{IEEEtran}
%\bibliography{graphonbiblio}


\begin{IEEEbiography}%
[{\includegraphics[width=1in,height=1.25in,clip,keepaspectratio]{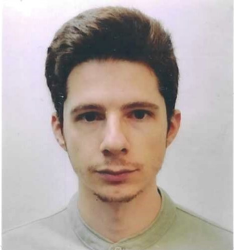}}]%
{Raoul Prisant}  received the Master degree in applied mathematics from
Politecnico di Torino, Turin, Italy, in 2023. 
He is currently a Ph.D.\ student in GIPSA-Lab, Grenoble, France. His research interests  include multi-agent systems, which was already the focus of his Bachelor and Master thesis,  and which has become the topic of his Ph.D., specializing in dynamics on large network systems. 
\end{IEEEbiography}

\begin{IEEEbiography}%
[{\includegraphics[width=1in,height=1.25in,clip,keepaspectratio]{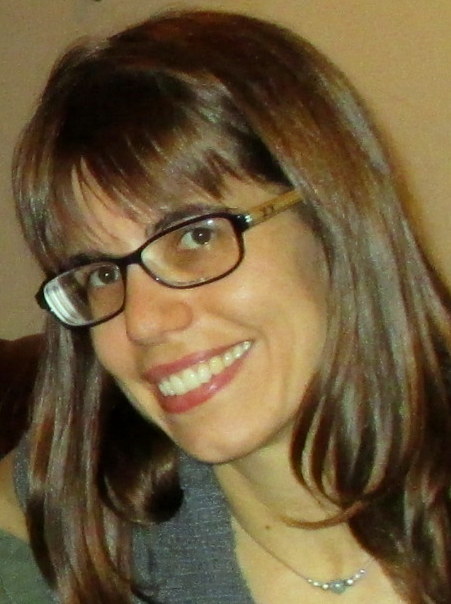}}]%
{Federica Garin} (Member, IEEE)  received the B.S.,
M.S., and Ph.D. degrees in applied mathematics from
Politecnico di Torino, Turin, Italy, in 2002, 2004, and
2008, respectively. Since 2010, she has been an Inria Researcher
with GIPSA-Lab, Grenoble France,
where she is currently serving as head of the Automatic Control and Diagnostics Department.
In 2008 and 2009,
she was a Postdoctoral Researcher with Università di
Padova, Padua, Italy. %, and in 2010, with Inria Grenoble, Montbonnot-Saint-Martin, France. 
She is Associate Editor of IEEE Control Systems Letters and Secretary of the European Control Association (EUCA).
Her current research
interests include  the theory of
networks and control systems, in particular large-scale network systems.
\end{IEEEbiography}

\begin{IEEEbiography}[{\includegraphics[width=1in,height =1.25in,clip,keepaspectratio]{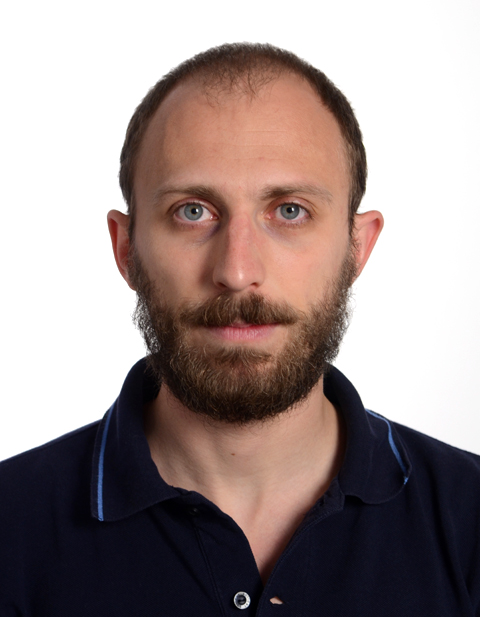}}]{Paolo Frasca} (Senior Member, IEEE) received the Ph.D. degree from the Politecnico di Torino, Turin, Italy, in 2009. %After post-doctoral appointments at IAC-CNR, Rome, and at Politecnico di Torino, Turin, 
He was an Assistant Professor at the University of Twente, Enschede, Netherlands, from 2013 to 2016. Since October 2016, he has been a CNRS Researcher at GIPSA-Lab, Grenoble, France, where since 2021 he has been leading the Dynamics and Control of Networks (DANCE) group. His research interests include control systems and network theory, with applications in transportation and social networks. Dr.\ Frasca has served as an Associate Editor for several conferences and journals, including the IEEE Control Systems Letters and Automatica.
\end{IEEEbiography}

\end{document}